\documentclass[%
pre,
twocolumn,
superscriptaddress,
]{revtex4-1}

\usepackage{times}
\usepackage{subcaption}
\usepackage{ragged2e}
\DeclareCaptionJustification{justified}{\justifying}
\usepackage{algorithm}
\usepackage{algpseudocode}
\usepackage{amsmath}
\usepackage{amssymb}
\usepackage{amsthm}
\usepackage{bbm}
\usepackage{enumerate}
\usepackage{enumitem}
\usepackage{pgfplots, pgfplotstable}
\usepackage{placeins}
\usepackage[titletoc,title]{appendix}
\usepackage{wrapfig,lipsum}
\usepackage{graphicx}
\usepackage{dcolumn}
\usepackage{bm}
\usepackage{overpic}
\usepackage{multirow}
\usepackage{tabularx,booktabs}
\usepackage{float}
\floatstyle{plaintop}
\restylefloat{table}

\algrenewcommand\algorithmicindent{1.0em}%
\makeatletter
\newcommand{\algmargin}{\the\ALG@thistlm}   
\makeatother
\algnewcommand{\parState}[1]{\State%
\parbox[t]{\dimexpr\linewidth-\algmargin}{\strut #1\strut}}

\newcolumntype{Y}{>{\centering\arraybackslash}X}

\newcommand{\Th}{$T_{\tiny \mbox{h}}$}
\newcommand{\Tl}{$T_{\tiny \mbox{l}}$}

\begin{document}

\title{Physics-Inspired Optimization for Quadratic Unconstrained Problems Using a Digital Annealer}

\author{Maliheh Aramon, Gili Rosenberg, Elisabetta Valiante}
\affiliation{1QB Information Technologies Inc. (1QBit), 458-550 Burrard Street, Vancouver, BC V6C 2B5, Canada}

\author{Toshiyuki Miyazawa, Hirotaka Tamura}
\affiliation{Fujitsu Laboratories Ltd., 4-1-1 Kamikodanaka, Nakahara-ku, Kawasaki, Kanagawa 211-8588, Japan}

\author{Helmut G. Katzgraber}
\affiliation{Microsoft Quantum, Microsoft, Redmond, WA 98052, USA}
\affiliation{Department of Physics and Astronomy, Texas A\&M University, College Station, TX 77843-4242, USA}
\affiliation{Santa Fe Institute, 1399 Hyde Park Road, Santa Fe, NM 87501, USA}

\pagestyle{empty}
\date{\today}

\begin{abstract}
The Fujitsu Digital Annealer is designed to solve fully connected
quadratic unconstrained binary optimization (QUBO) problems. It is
implemented on application-specific CMOS hardware and currently solves
problems of up to 1024 variables. The Digital Annealer's algorithm
is currently based on simulated annealing; however, it
differs from it in its utilization of an efficient parallel-trial scheme
and a dynamic escape mechanism. In addition, the Digital Annealer
exploits the massive parallelization that custom
application-specific CMOS hardware allows. We compare the performance of
the Digital Annealer to simulated annealing and parallel tempering with
isoenergetic cluster moves on two-dimensional and fully connected
spin-glass problems with bimodal and Gaussian couplings. These represent
the respective limits of sparse versus dense problems, as well as
high-degeneracy versus low-degeneracy problems. Our results show that
the Digital Annealer currently exhibits a time-to-solution speedup of
roughly two orders of magnitude for fully connected spin-glass problems
with bimodal or Gaussian couplings, over the single-core implementations
of simulated annealing and parallel tempering Monte Carlo used in this
study. The Digital Annealer does not appear to exhibit a speedup for
sparse two-dimensional spin-glass problems, which we explain on
theoretical grounds. We also benchmarked an early implementation of the
Parallel Tempering Digital Annealer. Our results suggest an improved scaling over the
other algorithms for fully connected problems of average difficulty with bimodal disorder. 
The next generation of the Digital Annealer is expected to be able to solve fully connected problems 
up to 8192 variables in size. This would enable the study of fundamental physics problems and industrial
applications that were previously inaccessible using standard computing hardware or 
special-purpose quantum annealing machines.
\end{abstract}


\maketitle

\section{Introduction}
\label{sec:intro}

Discrete optimization problems have ubiquitous applications in various
fields and, in particular, many NP-hard combinatorial optimization
problems can be mapped to a quadratic Ising model~\cite{lucas:14} or,
equivalently, to a quadratic unconstrained binary optimization (QUBO)
problem. Such problems arise naturally in many fields of research,
including finance~\cite{rosenberg:16}, chemistry~\cite{hernandez:16x,hernandez:17}, 
biology~\cite{perdomo:12,li:18},
logistics and scheduling~\cite{venturelli:15b,neukart:17}, and machine
learning~\cite{crawford:16x,khoshaman:19,henderson:18x,levit:17}.  
For this reason, there is much interest in solving these
problems efficiently, both in academia and in industry.

The impending end of Moore's law~\cite{moore:65} signals that relying on
traditional silicon-based computer devices is not expected to sustain
the current computational performance growth rate. In light of this,
interest in novel computational technologies has been steadily
increasing. The introduction of a special-purpose quantum annealer by
D-Wave Systems Inc.~\cite{johnson:11} was an effort in this direction,
aimed at revolutionizing how computationally intensive
discrete optimization problems are solved using quantum fluctuations.

Despite continued efforts to search for a scaling advantage of quantum
annealers over algorithms on conventional off-the-shelf CMOS hardware,
there is as yet no consensus.  Efforts to benchmark quantum annealers
against classical counterparts such as simulated annealing (SA)~\cite{kirkpatrick:83} have abounded~\cite{johnson:11,dickson:13,boixo:14,katzgraber:14,ronnow:14a,katzgraber:15,heim:15,hen:15a,albash:15a,martin-mayor:15a,marshall:16,denchev:16,king:17,albash:18,mandra:18,mandra:16b}.
Although for some classes of synthetic problems a large speedup was
initially found, those problems were subsequently shown to have a
trivial logical structure, such that they can be solved more efficiently
by more-powerful classical algorithms~\cite{mandra:17a}. To the best of
our knowledge, the only known case of speedup is a constant speedup for
a class of synthetic problems~\cite{mandra:18} and, so far, there is no
evidence of speedup for an industrial application. The hope is that
future improvements to the quantum annealer and, in particular, to its
currently sparse connectivity and low precision due to analog noise,
will demonstrate the power of quantum effects in solving optimization
problems~\cite{hamerly:18x,katzgraber:18}. With the same goal in mind,
researchers have been inspired to push the envelope for such problems on
novel hardware, such as the coherent Ising machine~\cite{hamerly:18x}, as
well as on graphics processing units (GPU)~\cite{king:17,albash:18} and
application-specific CMOS hardware~\cite{matsubara:17,tsukamoto:17}.
Similarly, efforts to emulate quantum effects in classical algorithms---often 
referred to as quantum- or physics-inspired methods---run on
off-the-shelf CMOS hardware have resulted in sizable advances in the
optimization domain (see, e.g., Ref.~\cite{mandra:16b} for an
overview of different algorithms).

Fujitsu Laboratories has recently developed application-specific CMOS
hardware designed to solve fully connected QUBO problems (i.e., on
complete graphs), known as the {\it{Digital Annealer}}~(DA)~\cite{matsubara:17, tsukamoto:17}.  The DA hardware is currently able to
treat Ising-type optimization problems of a size up to $1024$ variables,
with $26$ and $16$ bits of (fixed) precision for the biases and variable
couplers, respectively. The DA's algorithm, which we refer to as ``the
DA'', is based on simulated annealing, but differs in several ways (see
Sec.~\ref{sec:algorithms}), as well as in its ability to take
advantage of the massive parallelization possible when using a custom,
application-specific CMOS hardware. Previous efforts of running simulated annealing 
in parallel include executing different iterations in parallel on an AP1000 massively parallel 
distributed-memory multiprocessor \cite{sohn:95,sohn:96}. In addition to the DA, a version of the
Digital Annealer, which we refer to as ``the PTDA'', and which uses
parallel tempering Monte Carlo~\cite{swendsen:86,geyer:91,hukushima:96,earl:05,katzgraber:06a} for the
algorithmic engine is now available. In particular, it has been shown
that physics-inspired optimization techniques such as simulated
annealing and parallel tempering Monte Carlo typically outperform
specialized quantum hardware \cite{mandra:16b} such as the D-Wave
devices. 

Much of the benchmarking effort has centred around spin glasses, a type
of constraint satisfaction problem, in part due to their being the
simplest of the hard Boolean optimization problems.  Furthermore,
application-based benchmarks from, for example, industry tend to be structured
and, therefore, systematic benchmarking is difficult.  As such, spin glasses have
been used extensively to benchmark algorithms on off-the-shelf CPUs
\cite{wang:15e,wang:15}, novel computing technologies such as quantum
annealers \cite{karimi:17a,katzgraber:14,katzgraber:15,venturelli:15a},
and coherent Ising machines \cite{hamerly:18x}.  In this paper, we
benchmark the DA and the PTDA on spin-glass problems, comparing them to
simulated annealing \cite{isakov:15} and parallel tempering Monte Carlo
with isoenergetic cluster moves \cite{zhu:15b,zhu:16y} (a variant of
Houdayer cluster updates \cite{houdayer:01} within the context of
optimization and not the thermal simulation of spin-glass systems), both
state-of-the-art, physics-inspired optimization algorithms. For other 
alternative classical optimization techniques used in the literature to solve 
QUBO problems, the interested reader is referred to \cite{hen:15a,rosenberg:16a,mandra:16b}.

The paper is organized as follows. Section~\ref{sec:algorithms}
describes the algorithms we have benchmarked. In
Sec.~\ref{sec:parallel_vs_single} we probe the advantage of
parallel-trial over single-trial Monte Carlo moves and in
Sec.~\ref{sec:scaling} we discuss the methodology we have used for
measuring time to solution. In Sec.~\ref{sec:benchmark_problems} we
introduce the problems benchmarked. The experimental results are
presented and discussed in Sec.~\ref{sec:results}.  Finally, our
conclusions are presented in Sec.~\ref{sec:conclusions}.  The
parameters used for our benchmarking are given in \ref{sec:appendix_parameters}.

\section{Algorithms}
\label{sec:algorithms}
In this paper, we compare several Monte Carlo (MC) algorithms and their
use for solving optimization problems.

\subsection{Simulated Annealing}
Simulated annealing (SA) \cite{kirkpatrick:83} is a generic algorithm
with a wide application scope. The SA algorithm starts
from a random initial state at a high temperature.  Monte Carlo updates
at decreasing temperatures are then performed. Note that the
temperatures used follow a predefined schedule. 
\begin{algorithm}[H]
\caption{ \footnotesize{Simulated Annealing (SA)}}
\label{algo:SA}
\footnotesize
\begin{algorithmic}[1]
\For  {each run}
	\State {initialize to random initial state}
	\For {each temperature}
		\For  {each MC sweep at this temperature}
			\For {each variable} 
				\State {propose a flip}
				\State {if accepted, update the state and effective fields}
			\EndFor
		\EndFor
	\State {update the temperature}
	\EndFor
\EndFor
\end{algorithmic}
\end{algorithm}
When the simulation stops, one expects to find a low-temperature state, ideally the global
optimum (see Algorithm~\ref{algo:SA} for details).  The high-temperature
limit promotes diversification, whereas the low-temperature limit
promotes intensification. To increase the probability of finding the
optimum, this process is repeated multiple times (referred to as
``runs''), returning the best state found. The computational complexity
of each Monte Carlo sweep in SA is $\mathcal{O}(N^2)$ for
fully connected problems with $N$ variables, because each sweep includes
$N$ update proposals, and each accepted move requires updating $N$
effective fields, at a constant computational cost.

\subsection{The Digital Annealer's Algorithm}

The DA's algorithmic engine \cite{tsukamoto:17,matsubara:17} is based on
SA, but differs from it in three main ways (see Algorithm
\ref{algo:DA}). First, it starts all runs from the same arbitrary state,
instead of starting each run from a random state. This results in a
small speedup due to its avoiding the calculation of the initial $N$
effective fields and the initial energy for each run. Second, it uses a
{\em parallel-trial} scheme in which each Monte Carlo step considers a
flip of all variables (separately), in parallel. If at least one flip is
accepted, one of the accepted flips is chosen uniformly at random and it
is applied. Recall that in SA, each Monte Carlo step considers a flip of
a single variable only (i.e., {\em single trial}). The advantage of the
parallel-trial scheme is that it can boost the acceptance probability,
because the likelihood of accepting a flip out of $N$ flips is typically
much higher than the likelihood of flipping a particular variable (see
Sec.~\ref{sec:parallel_vs_single}). Parallel rejection algorithms on
GPU \cite{Niemi:11,ferrero:14} are examples of similar efforts in the literature to 
address the low acceptance probability problem in Monte Carlo methods. 
Finally, the DA employs an escape mechanism called a {\em dynamic offset}, such that if no flip was
accepted, the subsequent acceptance probabilities are artificially
increased by subtracting a positive value from the difference in energy
associated with a proposed move. This can help the algorithm to
surmount short, narrow barriers.

\begin{algorithm}[H]
\caption{\footnotesize{The Digital Annealer's Algorithm}}
\label{algo:DA}
\footnotesize
\begin{algorithmic}[1]
\State {initial\_state $\gets$ an arbitrary state}
\For {each run}
	\State {initialize to initial\_state}
	\State {$E_{\mbox{offset}} \gets 0$}
	\For  {each MC step (iteration)}
		\State {if due for temperature update, update the temperature}
		\For {each variable $j$, in parallel} 
			\State {propose a flip using $\Delta E_j - E_{\mbox{offset}}$}
			\State {if accepted, record}
		\EndFor
		\If {at least one flip accepted}
			\State {choose one flip uniformly at random amongst them}
			\State {update the state and effective fields, in parallel}
			\State {$E_{\mbox{offset}} \gets 0$}
		\Else 
			\State{$E_{\mbox{offset}} \gets E_{\mbox{offset}} + \mbox{offset\_increase\_rate}$}
		\EndIf
	\EndFor
\EndFor\end{algorithmic}
\end{algorithm}

Furthermore, the application-specific CMOS hardware allows for massive
parallelization that can be exploited for solving optimization problems
faster. For example, in the DA, evaluating a flip of all variables is
performed in parallel, and when a flip is accepted and applied, the
effective fields of all neighbours are updated in parallel. Note that
this step requires a constant time, regardless of the number of
neighbours, due to the parallelization on the hardware, whereas the
computational time of the same step in SA increases linearly in the
number of neighbours.

In order to understand the logic behind the DA, it is helpful to
understand several architectural considerations that are specific to the
DA hardware. In the DA, each Monte Carlo step takes the same
amount of time, regardless of whether a flip was accepted (and therefore
applied) or not. In contrast, in a CPU implementation of SA, accepted
moves are typically much more computationally costly than rejected moves,
that is, $\big[\mathcal{O}(N)$ vs. $\mathcal{O}(1)\big]$, due to the need to
update $N$ effective fields versus none if the flip is rejected. As a
result, in the DA, the potential boost in acceptance probabilities (from using the
parallel-trial scheme) is highly desirable. In addition, 
in the DA, the computational complexity of updating the
effective fields is constant regardless of the connectivity of the
graph. Comparing this with SA, the computational complexity of updating
the effective fields is $\mathcal{O}(N)$ for fully connected graphs, but
it is $\mathcal{O}(d)$ for fixed-degree graphs (in which each node has
$d$ neighbours). Therefore, running SA on a sparse graph is typically
faster than on a dense graph, but the time is the same for the DA. For
this reason, it is expected that the speedup of the DA over SA be, in
general, higher for dense graphs than for sparse ones.

Finally, it is worth mentioning that the pseudorandom number generator algorithm implemented 
in the hardware is similar to a twisted generalized feedback shift register algorithm and gives a 
sufficiently long period of $2^{19937}$ \cite{matsumoto:92}. 

\subsection{Parallel Tempering with Isoenergetic Cluster Moves}
In parallel tempering (PT)
\cite{swendsen:86,geyer:91,hukushima:96,earl:05,katzgraber:06a,katzgraber:09e}
(also known as replica-exchange Monte Carlo), multiple replicas of the
system are simulated at different temperatures, with periodic exchanges
based on a Metropolis criterion between neighbouring temperatures. Each
replica, therefore, performs a random walk in temperature space,
allowing it to overcome energy barriers by temporarily moving to a
higher temperature. The higher-temperature replicas are typically at a
high enough temperature that they inject new random states into the
system, essentially re-seeding the algorithm continuously, obviating (at
least partially) the need to perform multiple runs. PT has been used
effectively in multiple research fields \cite{earl:05}, and often
performs better than SA, due to the increased mixing.

The addition of isoenergetic cluster moves (ICM)
\cite{zhu:15b,houdayer:01} to PT, which flip multiple variables at a
time, can allow for better exploration of the phase space, but only if
the variable clusters do not span the whole system
\cite{zhu:16y,zhu:15b}. ICM is a generalization of Houdayer's cluster
algorithm, which was tailored for two-dimensional spin-glass problems
\cite{houdayer:01}. To perform ICM, two copies (or more) of the system
are simulated at the same temperature. The states of those two replicas
are then compared, to find a cluster of variables (i.e., a connected component)
that are opposite. In the case of QUBO problems, opposite variables are defined as
having a product of zero.  Finally, the move is applied by swapping the
states of the opposite variables in the two replicas. The total energy
of the two replicas is unchanged by this move, such that it is rejection free.
The combination of PT and
ICM, PT+ICM (also known as {\em borealis}; see
Algorithm~\ref{algo:PTICM} \cite{zhu:16y}), has been shown to be highly
effective for low-dimensionality spin-glass-like problems
\cite{katzgraber:15,zhu:16,mandra:16b}, but it does not provide a benefit
for problems defined on fully connected graphs. This can be understood
by noting that when the clusters span the system, ICM essentially
results in swapping the states completely.

\begin{algorithm}[H]
\caption{\footnotesize{Parallel Tempering with Isoenergetic Cluster Moves (PT+ICM)}}
\label{algo:PTICM}
\footnotesize
\begin{algorithmic}[1]
\State {initialize all replicas with random initial states}
\For  {each MC sweep}
	\For {each replica, for each variable} 
		\State {propose a flip}
		\State {if accepted, update the state and effective fields}
	\EndFor
	\For {each pair of sequential replicas}
		\State {propose a replica exchange}
		\State {if accepted, swap the temperatures between the replicas}
	\EndFor
	\State {perform ICM update, swapping the states of a cluster of variables that have opposite states in the two replicas; update the states and the effective fields for both replicas}
\EndFor
\end{algorithmic}
\end{algorithm}

\subsection{The Parallel Tempering Digital Annealer's Algorithm}

Because the DA's algorithm is based on SA, and given the often superior
results that PT gives over SA (see, e.g., Ref.~\cite{mandra:16b}),
Fujitsu Laboratories has also developed a Parallel Tempering Digital
Annealer (PTDA). We had access to an early implementation of a
PTDA. In the PTDA, the sweeps in each replica are performed as in the
DA, including the parallel-trial scheme, parallel updates, and using the
dynamic offset mechanism, but the PT moves are performed on a CPU. The
temperatures are set automatically based on an adaptive scheme by
Hukushima {et~al.~\cite{hukushima:99}. In this scheme, the high and
low temperatures are fixed, and intermediate temperatures are adjusted
with the objective of achieving an equal replica-exchange probability
for all adjacent temperatures. Having equal replica-exchange acceptance
probabilities is a common target, although other schemes exist
\cite{katzgraber:06a}.

The next generation of the Digital Annealer is expected to be able to
simulate problems on complete graphs up to $8192$ variables in size,
to have faster annealing times, and to perform the replica-exchange
moves on the hardware, rather than on a CPU. This is significant,
because when performing a computation in parallel, if a portion of the
work is performed sequentially, it introduces a bottleneck that
eventually dominates the overall run time (as the number of parallel
threads is increased).  Amdahl's Law \cite{amdahl:67} quantifies this by
stating that if the sequential part is a fraction $\alpha$ of the total
work, the speedup is limited to $1/\alpha$ asymptotically.

\section{Parallel-trial versus Single-trial Monte Carlo}
\label{sec:parallel_vs_single}

To illustrate the advantage of parallel-trial Monte Carlo updates as implemented in the DA
over single-trial Monte Carlo updates, let us calculate their respective
acceptance probabilities.  The acceptance probability for a particular
Monte Carlo move is given by the Metropolis criterion
${\mathcal{A}(\Delta E_i, T) \equiv e^{-\Delta E_i / T}}$, where $\Delta
E_i$ denotes the difference in energy associated with flipping variable
$i$, and $T$ is the temperature. The single-trial acceptance probability
is then given by
\begin{align}
\label{eq:single_trial}
\mathcal{P}_{\tiny \mbox{s}}(T) = \frac1{N} \sum_i \mathcal{A}(\Delta E_i, T) \,,
\end{align}
where $N$ is the number of variables.  In contrast, the parallel-trial
acceptance probability is given by the complement probability of not
accepting a move,
\begin{align}
\label{eq:parallel_trial}
\mathcal{P}_{\tiny \mbox{p}}(T) = 1 - \prod_i \left[ 1-\mathcal{A}(\Delta E_i, T) \right].
\end{align}
At low temperatures, we expect the acceptance probability to reach zero,
in general. In the limit ${\mathcal{A} \to 0}$, a first-order
approximation of the parallel-trial acceptance probability gives
\begin{align}
\label{eq:parallel_trial_approx}
\mathcal{P}_{\tiny \mbox{p}}(T) \simeq \sum_i \mathcal{A}(\Delta E_i, T) \,.
\end{align}
This indicates that in the best case, there is a speedup by a factor of
$N$ at low temperatures. In contrast, at a high enough temperature, all
moves are accepted, hence ${\mathcal{A} \to 1}$. In this limit, it is
clear that both the single-trial and parallel-trial acceptance
probabilities reach $1$, so parallel-trial Monte Carlo does not have an
advantage over single-trial Monte Carlo.

To quantify the difference between parallel-trial and single-trial Monte
Carlo, we perform a Monte Carlo simulation at constant temperature for a
sufficiently large number of sweeps to reach thermalization. Once the
system has thermalized, we measure the single-trial and parallel-trial
acceptance probabilities at every move. This has been repeated for a number of
sweeps, and for multiple temperatures and multiple problems.

The results of such an experiment are presented in
Fig.~\ref{fig:single_vs_parallel}, for problems of size $N = 64$ of the
four problem classes described in detail in
Sec.~\ref{sec:benchmark_problems}. The problem classes include
two-dimensional (2D) and fully connected [Sherrington--Kirkpatrick (SK)]
spin-glass problems with bimodal and Gaussian disorder. The results for
all the problem classes except for the 2D-bimodal class follow the
expected pattern of the acceptance probabilities reaching zero at low
temperatures and one at high temperatures. In the 2D-bimodal case, there
is a huge ground-state degeneracy, such that even at the ground state
there are single variables for which a flip does not result in a change in
energy. This results in a positive single-trial acceptance probability
even at very low temperatures. For the same reason, the parallel-trial
probability reaches one even for very low temperatures.

\begin{figure}[!htbp]
\begin{subfigure}[b]{0.5\linewidth}
\centering  
\begin{overpic}[width=1\linewidth]{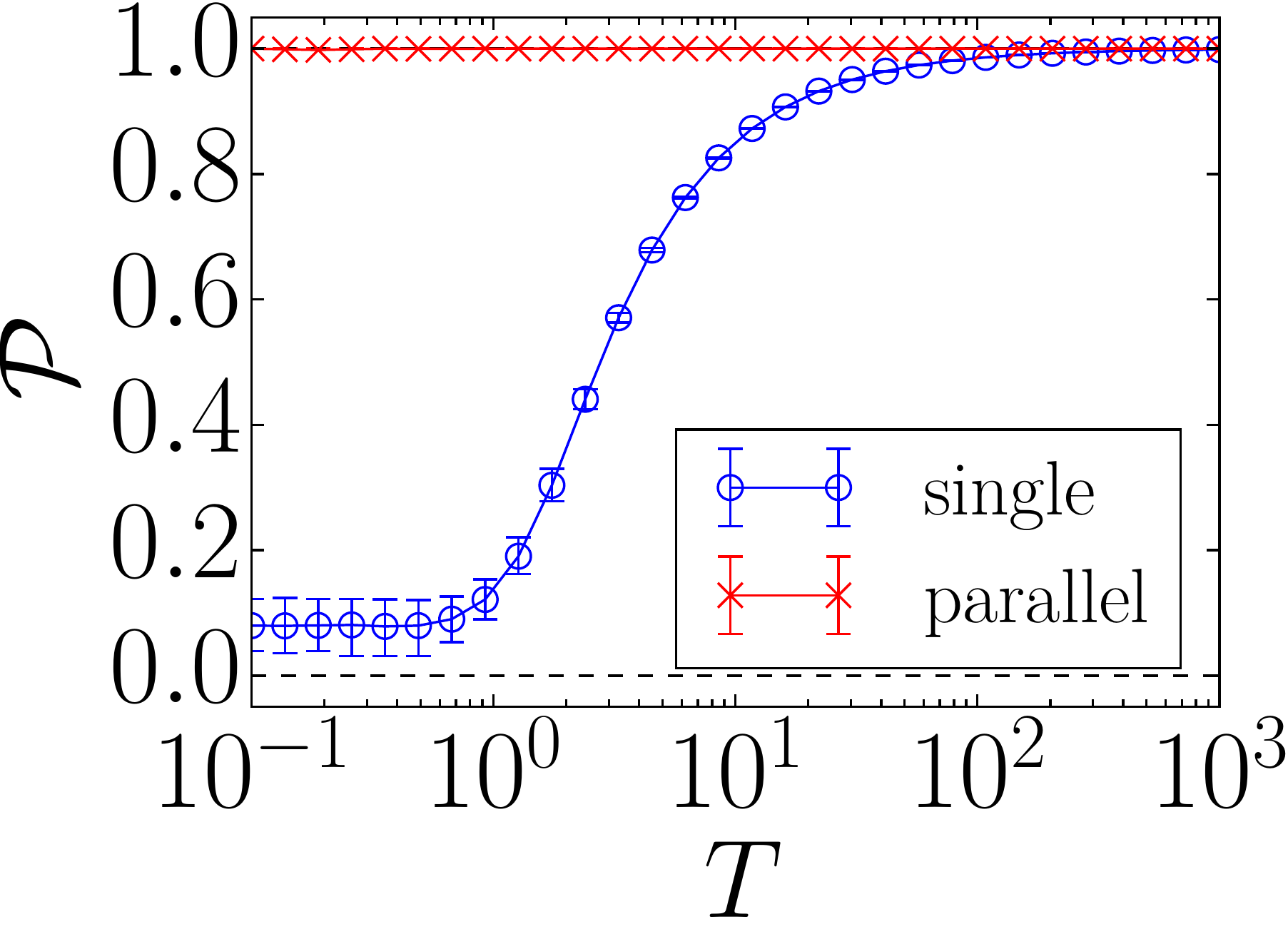}
\put(80,58){{(a)}}
\end{overpic}
\caption{}
\label{fig:single_vs_parallel_2d_bimodal}
\end{subfigure}
\begin{subfigure}[b]{0.5\linewidth}
\centering
\begin{overpic}[width=1\linewidth]{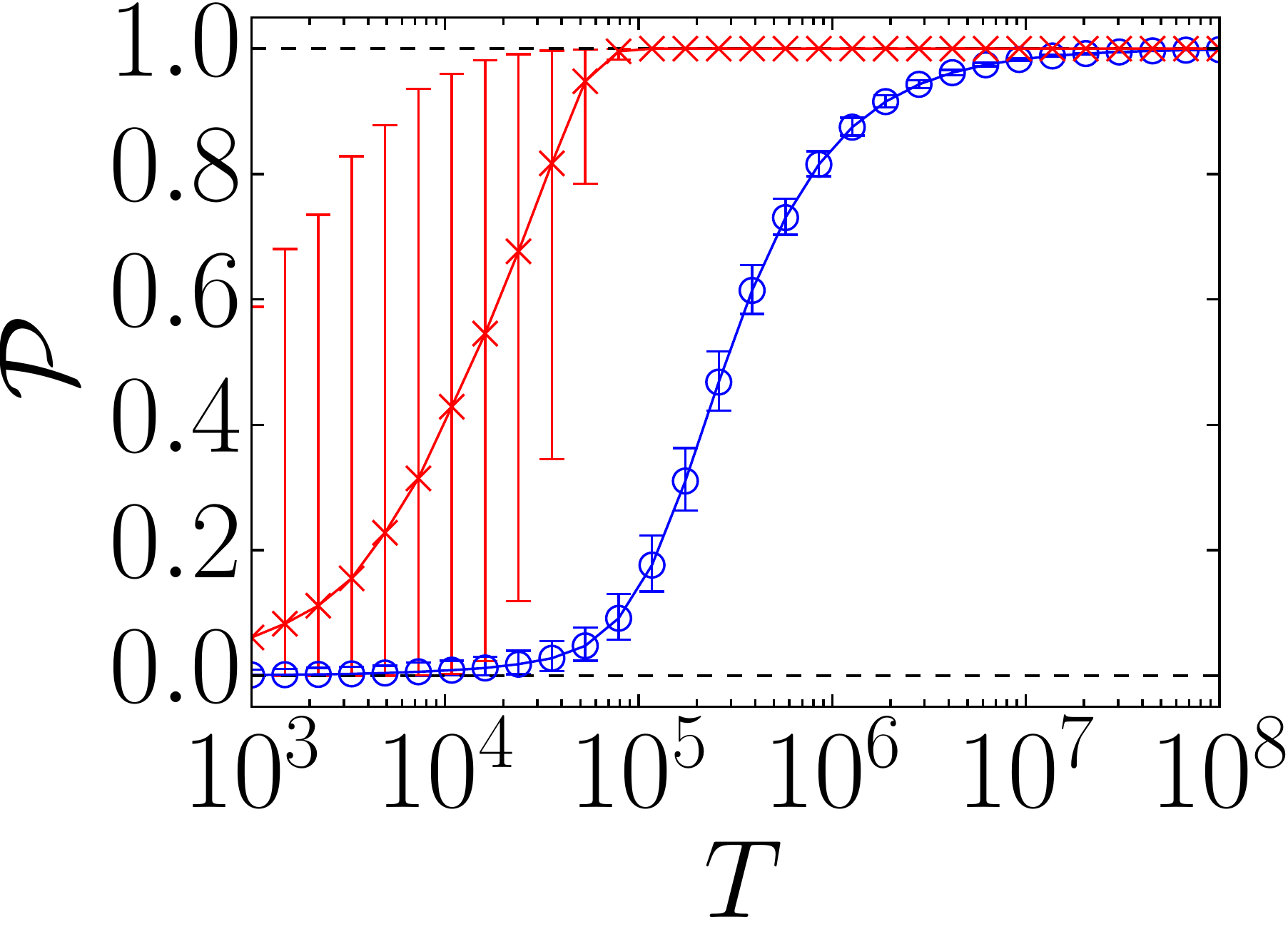}
\put(80,58){{(b)}}
\end{overpic}
\caption{}
\label{fig:single_vs_parallel_2d_gaussian}
\end{subfigure}
\begin{subfigure}[b]{0.5\linewidth}
\centering  
\begin{overpic}[width=1\linewidth]{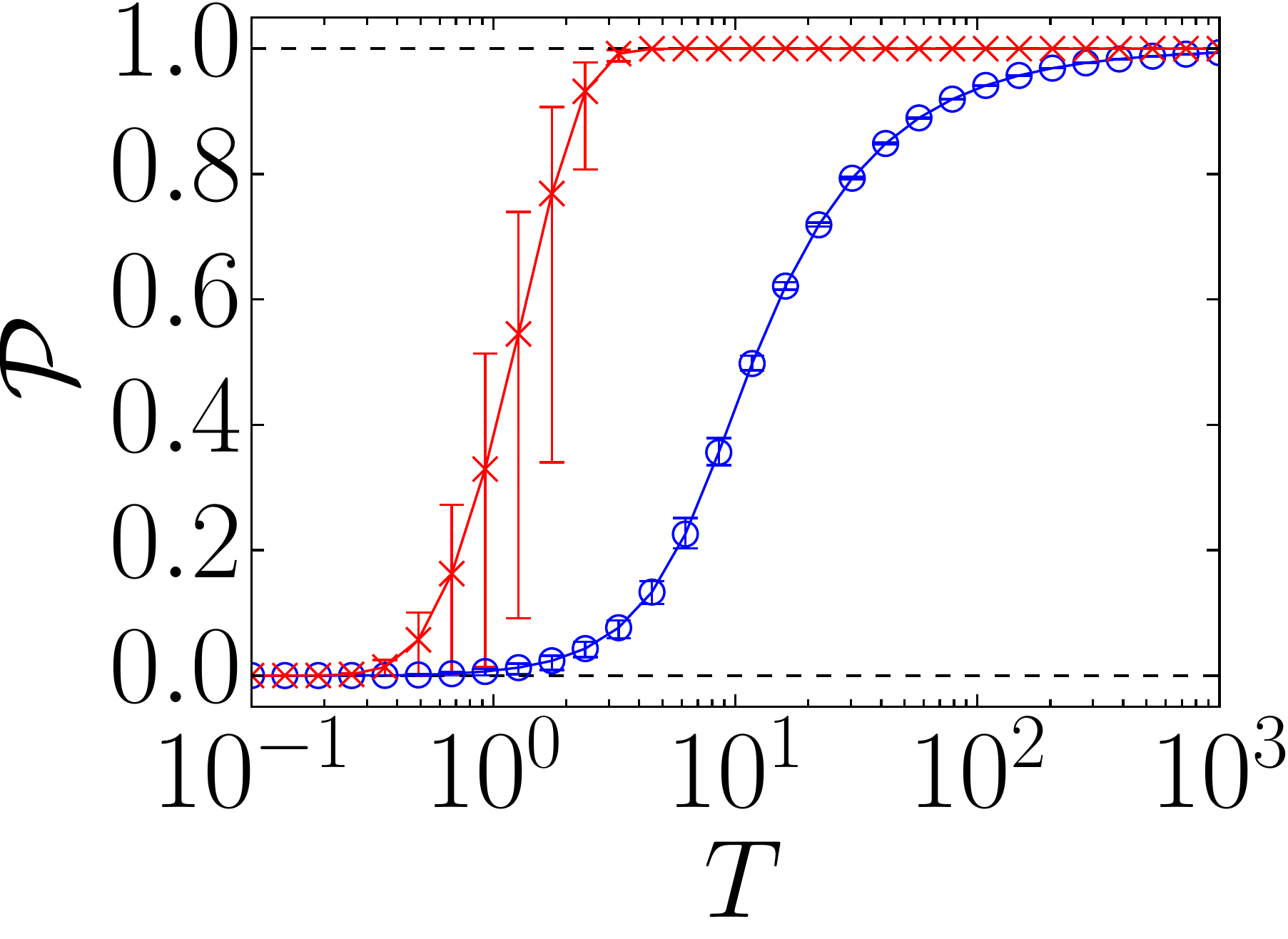}
\put(80,58){{(c)}}
\end{overpic}
\caption{}
\label{fig:single_vs_parallel_sk_bimodal}
\end{subfigure}
\begin{subfigure}[b]{0.5\linewidth}
\centering
\begin{overpic}[width=1\linewidth]{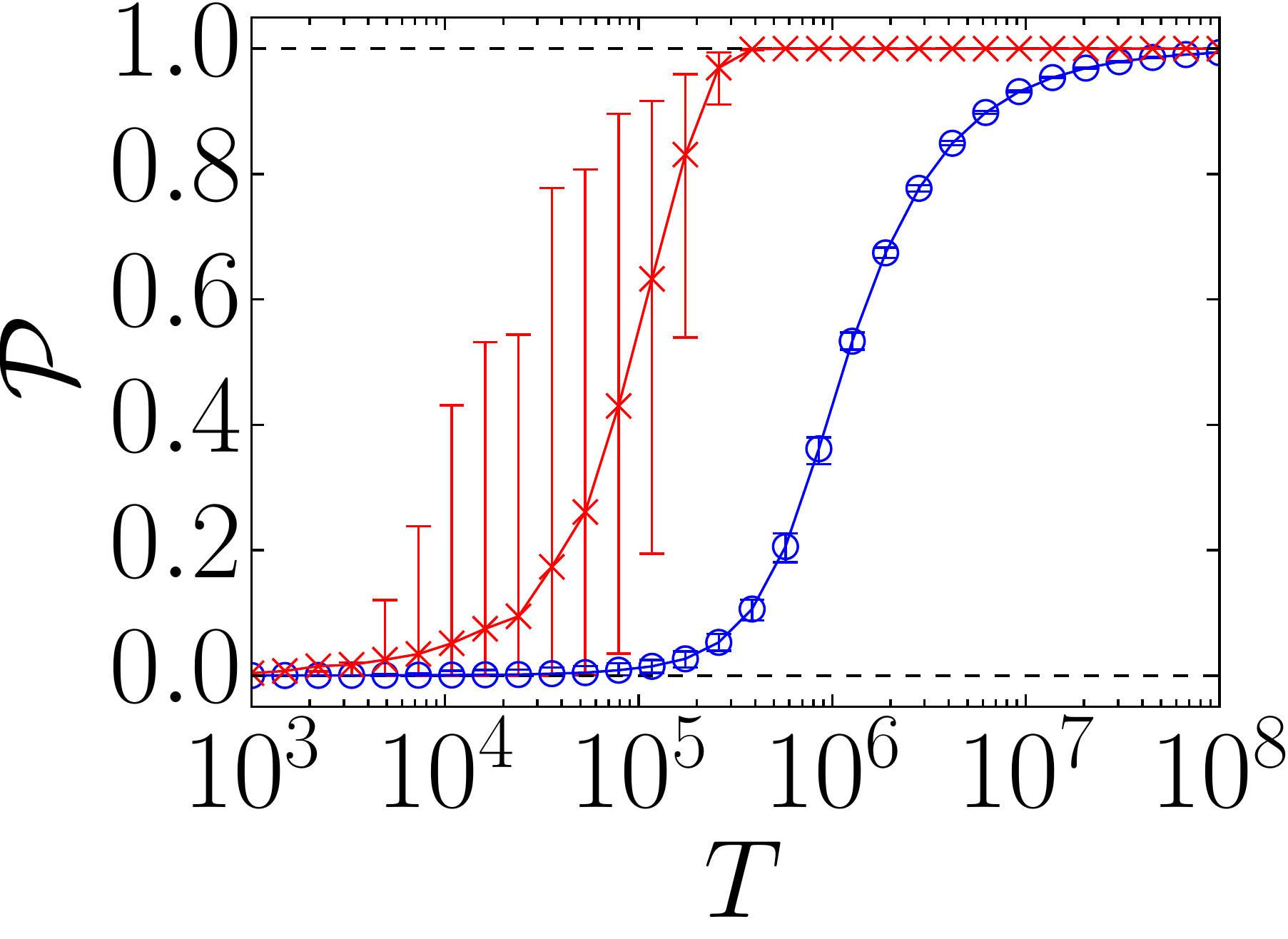}
\put(80,58){{(d)}}
\end{overpic}\caption{}
\label{fig:single_vs_parallel_sk_gaussian}
\end{subfigure}
\caption{
Mean single-trial and parallel-trial acceptance probabilities
$(\mathcal{P})$ vs. the temperature ($T$), for four problem
classes: (a) 2D-bimodal, (b) 2D-Gaussian, (c) SK-bimodal, and (d)~
SK-Gaussian. Error bars are given by the $5$th and $95$th percentiles of
the acceptance probabilities in the configuration space. A
constant-temperature Monte Carlo simulation has been run for $10^5$ sweeps to
thermalize, after which the acceptance probabilities are measured for $5
\cdot 10^3$ sweeps. For each problem class, the simulation has been performed
on $100$ problem instances of size $N = 64$ with $10$ repeats per
problem instance. The horizontal dashed lines show the minimum and
maximum attainable probabilities, zero and one, respectively. Acceptance
probabilities for the parallel-trial scheme approach unity as a function of
simulation time considerably faster than in the single-trial scheme.}
\label{fig:single_vs_parallel} 
\end{figure}

To quantify the acceptance probability advantage of parallel-trial over
single-trial updates, it is instructive to study the parallel-trial
acceptance probability divided by the single-trial acceptance
probability, as presented in Fig.~\ref{fig:single_vs_parallel_boost}.
For all problem classes except 2D-bimodal, the advantage at low
temperatures is indeed a factor of $N$, as suggested by dividing
Eq.~\eqref{eq:parallel_trial_approx} by Eq.~\eqref{eq:single_trial}. As
explained above, in the 2D-bimodal case the single-trial acceptance
probability is nonnegligible at low temperatures, leading to a reduced
advantage. It is noteworthy that the advantage of the parallel-trial
scheme is maximal at low temperatures, where the thermalization time is
longer. As such, the parallel-trial scheme provides an acceptance probability boost
where it is most needed.

\begin{figure}[!htbp]
\begin{subfigure}[b]{0.5\linewidth}
\centering  
\begin{overpic}[width=1\linewidth]{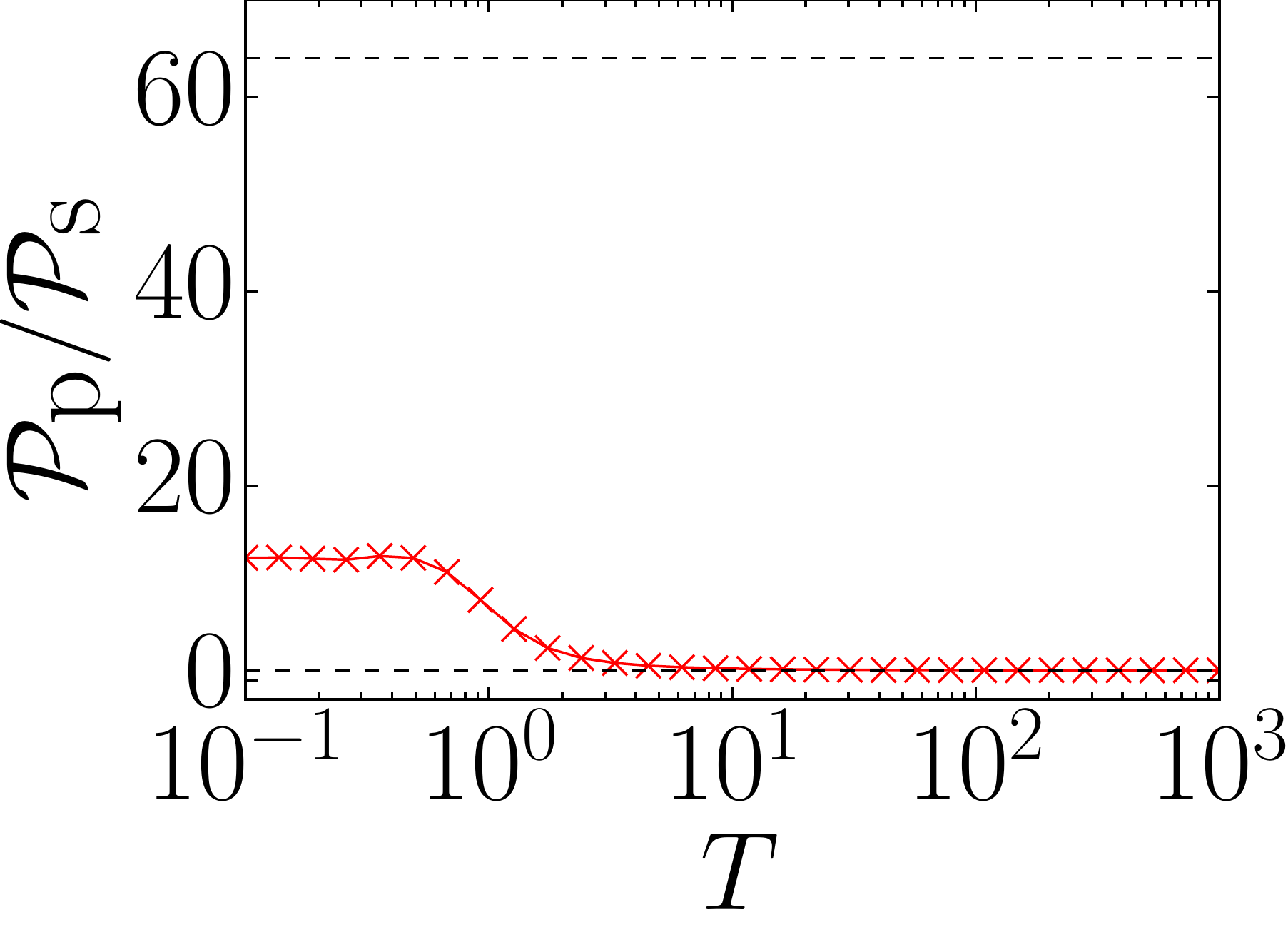}
\put(80,54){{(a)}}
\end{overpic}
\caption{}
\label{fig:single_vs_parallel_2d_bimodal_boost}
\end{subfigure}
\begin{subfigure}[b]{0.5\linewidth}
\centering
\begin{overpic}[width=1\linewidth]{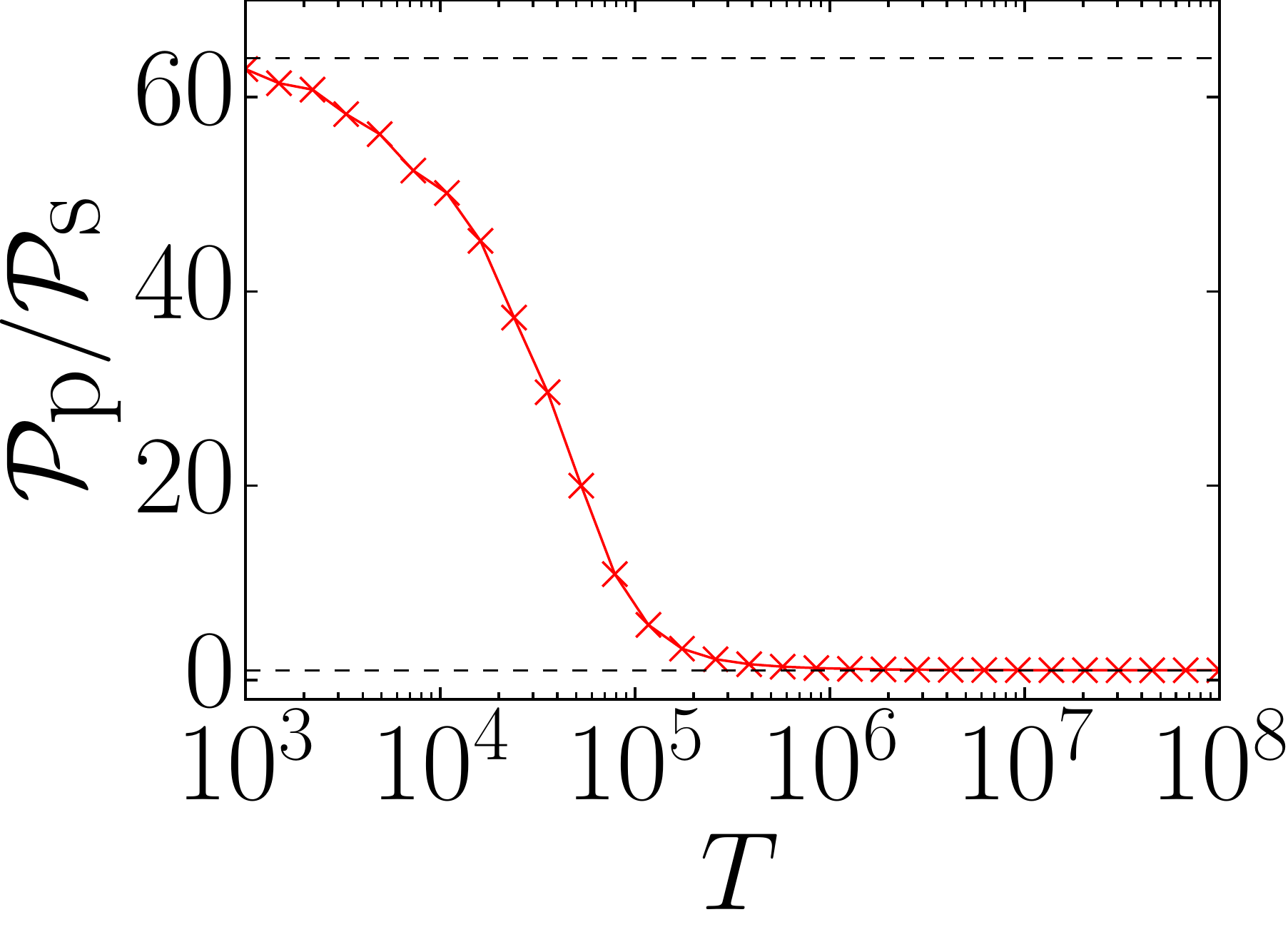}
\put(80,54){{(b)}}
\end{overpic}
\caption{}
\label{fig:single_vs_parallel_2d_gaussian_boost}
\end{subfigure}
\begin{subfigure}[b]{0.5\linewidth}
\centering  
\begin{overpic}[width=1\linewidth]{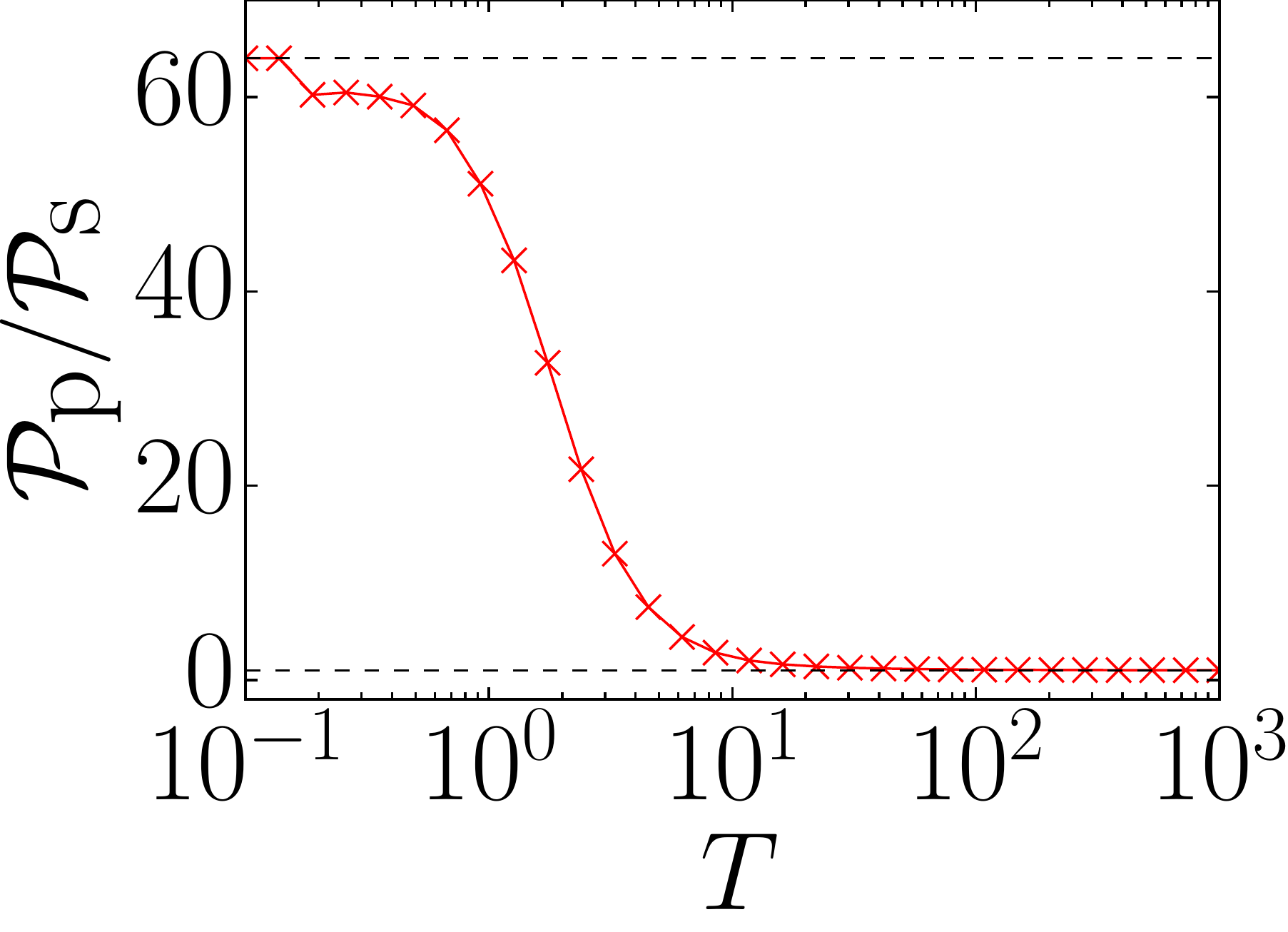}
\put(80,54){{(c)}}
\end{overpic}
\caption{}
\label{fig:single_vs_parallel_sk_bimodal_boost}
\end{subfigure}
\begin{subfigure}[b]{0.5\linewidth}
\centering
\begin{overpic}[width=1\linewidth]{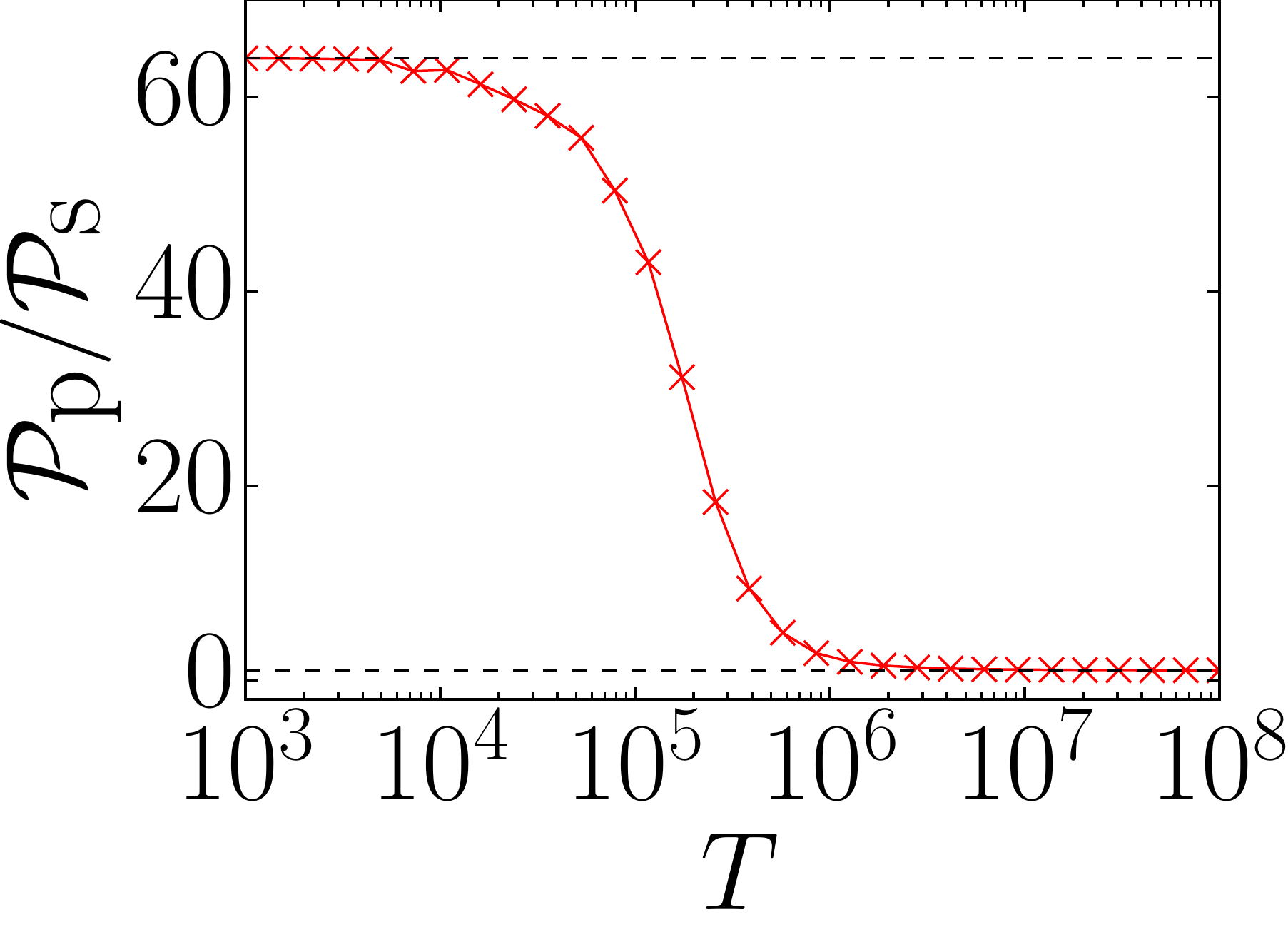}
\put(80,54){{(d)}}
\end{overpic}
\caption{}
\label{fig:single_vs_parallel_sk_gaussian_boost}
\end{subfigure}
\caption{
Ratio of the mean parallel-trial acceptance probability
$(\mathcal{P}_{\mbox{p}})$ to the mean single-trial acceptance probability
$(\mathcal{P}_{\mbox{s}})$ vs. the temperature $(T)$, for four
problem classes: (a) 2D-bimodal, (b) 2D-Gaussian, (c) SK-bimodal, and
(d) SK-Gaussian. A constant-temperature Monte Carlo simulation has been run
for $10^5$ sweeps to thermalize, after which the acceptance
probabilities are measured for $5 \cdot 10^3$ sweeps. For each problem
class, the simulation has been performed on $100$ problem instances of size $N
= 64$ with $10$ repeats per problem instance. The upper horizontal
dashed line indicates $N$, the number of variables, which is the
expected theoretical value of the ratio at low temperatures, as given by
Eq.~\eqref{eq:parallel_trial_approx} divided by
Eq.~\eqref{eq:single_trial}. The lower horizontal line indicates the
minimum value of the ratio, which is one. }
\label{fig:single_vs_parallel_boost}
\end{figure}

\section{Scaling Analysis}
\label{sec:scaling}

The primary objective of benchmarking is to quantify how the
computational effort in solving problems scales as the size of the
problem (e.g., the number of variables) increases. The algorithms we
consider here are all stochastic, and a common approach to measuring the
scaling of a probabilistic algorithm is to measure the total time
required for the algorithm to find a reference energy (cost) at least
once with a probability of $0.99$. The reference energy is represented
by the optimal energy if available or, otherwise, by the best known
energy. We denote this time to solution by ``TTS'', and explain how it
is calculated in the rest of this section.

We consider the successive runs of a probabilistic algorithm as being a
sequence of binary experiments that might succeed in returning the
reference energy with some probability. Let us formally define $X_1,
X_2, \ldots, X_r$ as a sequence of random, independent outcomes of $r$
runs (experiments), where $\mathbbm{P}(X_i=1)=\theta$ denotes the probability of
success, that is, of observing the reference energy at the $i$-th run.
Defining 
\begin{equation}
Y=\sum_{i=1}^{r} X_i
\end{equation}
as the number of successful observations
in $r$ runs, we have
\begin{equation}
\mathbbm{P}(Y = y|\theta, r) = {r \choose y} (1-\theta)^{r-y}\theta^y .
\end{equation}
That is, $Y$ has a binomial distribution with parameters $r$ and
$\theta$. We denote the number of runs required to find the reference
energy with a probability of $0.99$ as $\mbox{R}_{99}$, which equals $r$
such that $\mathbbm{P}(Y \geq 1|\theta, r) = 0.99$. It can be verified
that
\begin{equation}
\mbox{R}_{99} = \frac{\log (1-0.99)}{\log(1-\theta)}
\end{equation}
and, consequently, that 
\begin{equation}
\mbox{TTS} = \tau \mbox{R}_{99} , 
\end{equation}
where $\tau$ is the time it takes to run the algorithm once. Because the
probability of success $\theta$ is unknown, the challenge is in
estimating $\theta$.

Instead of using the sample success proportion as a point estimate for
$\theta$, we follow the Bayesian inference technique to estimate the
distribution of the probability of success for each problem instance
\cite{hen:15a}. Having distributions of the success probabilities would
be helpful in more accurately capturing the variance of different
statistics of the TTS. In the Bayesian inference framework, we start
with a guess on the distribution of $\theta$, known as a prior, and
update it based on the observations from consecutive runs of the
algorithm in order to obtain the posterior distribution. Since the
consecutive runs have a binomial distribution, the suitable choice of
prior is a beta distribution \cite{abramowitz:64} that is the conjugate
prior of the binomial distribution. This choice guarantees that the
posterior will also have a beta distribution.  The beta distribution
with parameters $\alpha=0.5$ and $\beta=0.5$ (the Jeffreys prior) is
chosen as a prior because it is invariant under reparameterization of the
space and it learns the most from the data \cite{clarke:94}.  Updating
the Jeffreys prior based on the observations from consecutive runs, the
posterior distribution, denoted by $\pi(\theta)$, is $\pi(\theta) \sim
\mbox{Beta} (\alpha + y, \beta + r - y)$, where $r$ is the total number
of runs and $y$ is the number of runs in which the reference energy is
found.

To estimate the TTS for the entire population of instances with similar
parameters, let us assume that there are $I$ instances with the same
number of variables. After finding the posterior distribution
$\pi_i(\theta)$ for each instance $i$ in set $\{1, 2, \ldots, I\}$, we
use bootstrapping to estimate the distribution of the $q$-th percentile
of the TTS. This procedure is described in Algorithm~\ref{TTS}.
\begin{algorithm}[H]
\caption{\footnotesize{Estimating the Distribution of the $q$-th Percentile of the TTS}}
\label{TTS}
\footnotesize
\begin{algorithmic}[1]
\State {fix the number of bootstrap resamples to $B$ ($B = 5000$)}
\For  {$b=1, \ldots, B$}
\State {sample $I$ instances with replacement}
\For {each sampled instance $j$} 
\State {sample a value, $p_{jb}$, from its posterior probability $\pi_j(\theta)$}
\State {calculate $\mbox{R}_{99, jb} = \log (1-0.99) / \log(1-p_{jb})$}
\EndFor
\State {find the $q$-th percentile of the set $\{\mbox{R}_{99,jb}\}$ and denote it by $\mbox{R}_{99,qb}$}
\EndFor
\State {consider the empirical distribution of $(\tau\mbox{R}_{99,q1},\ldots, \tau\mbox{R}_{99,qB})$ as an approximation of the true $\mbox{TTS}_{q}$ distribution}
\end{algorithmic}
\end{algorithm}
The procedure for deriving the TTS is slightly different for the DA and
the PTDA. The anneal time of the algorithmic engine of the DA is not a linear function of the number of runs for a given
number of sweeps. We therefore directly measure the anneal time for a
given number of iterations and a given number of runs where the latter
is equal to the $\mbox{R}_{99}$. Each iteration (Monte Carlo step) in
the DA represents one potential update and each Monte
Carlo sweep corresponds to $N$ iterations.

It is important to note that the correct scaling is only observed if the
parameters of the solver are tuned such that the TTS is minimized.
Otherwise, a suboptimal scaling might be observed and incorrect
conclusions could be made.  Recall that the TTS is the product of the
$\mbox{R}_{99}$ and the time taken per run $\tau$. Let us consider a
parameter that affects the computational effort taken, such as the
number of sweeps. Increasing the number of sweeps results in the
algorithm being more likely to find the reference solution, hence
resulting in a lower $\mbox{R}_{99}$. On the other hand, increasing the
number of sweeps also results in a longer runtime, increasing $\tau$.
For this reason, it is typical to find that the TTS reaches infinity for
a very low or very high number of sweeps, and the goal is to
experimentally find a number of sweeps at which the TTS is minimized.

\section{Benchmarking Problems}
\label{sec:benchmark_problems}

A quadratic Ising model can be represented by a Hamiltonian (i.e., cost
function) of the form
\begin{align}
\label{eq:ising}
\mathcal{H} = - \sum_{(i,j) \in E} J_{ij} s_i s_j - \sum_{i \in V} h_i s_i \,.
\end{align}
Here, $s_i \in \{-1, 1\}$ represent Boolean variables, and the problem
is encoded in the biases $h_i$ and couplers $J_{ij}$. The sums are over
the vertices $V$ and weighted edges $E$ of a graph $G = (V,E)$.  It can
be shown that the problem of finding a spin configuration $\{s_i\}$ that
minimizes $\mathcal{H}$, in general, is equivalent to the NP-hard
weighted max-cut problem~\cite{juenger:95,pardella:08,liers:10,juenger:01}.  Spin glasses defined
on nonplanar graphs fall into the NP-hard complexity class.  However,
for the special case of planar graphs, exact, polynomial-time methods
exist \cite{groetschel:87}.

The algorithmic engine of the Digital Annealer can optimize instances of QUBO problems
in which the variables $x_i$ take values from $\{0,1\}$ instead of
$\{-1,1\}$. To solve a quadratic Ising problem described by the
Hamiltonian represented in Eq.~\eqref{eq:ising}, we can transform it
into a QUBO problem by taking $s_i = 2x_i -1$.

In the following, we explain the spin-glass problems used for benchmarking. 
\begin{enumerate}

\item[]{{\em 2D-bimodal} -- Two-dimensional spin-glass problems on a
torus (periodic boundaries), where couplings are chosen according to a
bimodal distribution, that is, they take values from $\{-1, 1\}$ with
equal probability.}

\item[]{{\em 2D-Gaussian} -- Two-dimensional spin-glass problems where
couplings are chosen from a Gaussian distribution with a mean of zero
and a standard deviation of one, scaled by $10^5$.}

\item[]{{\em SK-bimodal} -- Spin-glass problems on a complete
graph---also known as Sherrington--Kirkpatrick (SK) spin-glass problems
\cite{sherrington:75}---where couplings are chosen according to a
bimodal distribution, that is, they take values from $\{-1, 1\}$ with
equal probability.}

\item[]{{\em SK-Gaussian} -- SK spin-glass problems where couplings are
chosen from a Gaussian distribution with a mean of zero and a standard
deviation of one, scaled by $10^5$.}

\end{enumerate}
In all the problems, the biases are zero. The coefficients of
the 2D-Gaussian and SK-Gaussian problems are beyond the precision limit of
the current DA. In order to solve these problems using the DA, we have
used a simple scheme to first scale the coefficients up to their maximum
limit and then round to the nearest integer values. The maximum values
for the linear and quadratic coefficients are given by the precision
limits of the current DA hardware, that is, $2^{25}-1$ and $2^{15}-1$,
respectively. The problem instances are not scaled when solving them using SA
or PT (PT+ICM).

Our benchmarking experiment has been parametrized by the number of variables.
Specifically, we have considered nine different problem sizes in each
problem category and generated $100$ random instances for each problem
size. We have used the instance generator provided by the University of
Cologne Spin Glass Server \cite{juenger:sg} to procure the 2D-bimodal
and 2D-Gaussian instances. SK instances with bimodal and Gaussian
disorder have been generated as described above. Each problem instance
has then been solved by different Monte Carlo algorithms. Optimal solutions to
the 2D-bimodal and 2D-Gaussian problems have been obtained by a branch-and-cut
\cite{liers:05} technique available via the Spin Glass Server
\cite{juenger:sg}. The SK problems are harder than the two-dimensional
problems and the server does not find the optimal solution within the 15-minute 
time limit. For the SK-bimodal and the SK-Gaussian problems with
$64$ variables, we have used a semidefinite branch-and-bound
technique through the Biq Mac Solver \cite{biqMac} and BiqCrunch 
\cite{biqCrunch} to find the optimal solutions, respectively. For
problems of a size greater than $64$, the solution obtained by PT with
a large number of sweeps ($5 \cdot 10^5$ sweeps) is considered a good
upper bound for the optimal solution. We refer to the optimal solution
(or its upper bound) as the {\em reference energy}.

\section{Results and Discussion}
\label{sec:results}

In this paper, we have used an implementation of the PT (and PT+ICM)
algorithm based on the work of Zhu {et~al.}~\cite{zhu:16y,zhu:15b}.
The DA and PTDA algorithms are run on Fujitsu's Digital Annealer
hardware. For the SA simulations we have used the highly optimized, open source code by Isakov {et~al.}~\cite{isakov:15}.

The DA solves only a fully connected problem where the coefficients of
the absent vertices and edges in the original problem graph are set to
zero. In our benchmarking study, we have included both two-dimensional and SK problems to 
represent the two cases of sparsity and full connectivity, respectively. Furthermore, we have 
considered both bimodal and Gaussian disorder in order to account for problems with high or 
low ground-state degeneracy. The bimodal disorder results in an energy landscape that has 
a large number of free variables with zero effective local fields. As a result, the degeneracy 
of the ground state increases exponentially in the number of free variables, making it easier for 
any classical optimization algorithm to reach a ground state. Problem instances that have 
Gaussian coefficients further challenge the DA, due to its current limitations in terms of precision.

In what follows, we discuss our benchmarking results, comparing the
performance of different algorithms using two-dimensional and SK
spin-glass problems, with bimodal and Gaussian disorder. We further investigate how 
problem density affects the DA's performance. The parameters
of the algorithms used in this benchmarking study are presented in \ref{sec:appendix_parameters}.

\subsection{2D Spin-Glass Problems}
\label{subsec:2d_results}

Figure~\ref{fig:tts_2d} illustrates the TTS results of the DA, SA, PT+ICM, and 
PTDA for 2D spin-glass problems with bimodal and Gaussian disorder. In
all TTS plots in this paper, points and error bars represent the mean
and the $5$th and $95$th percentiles of the TTS distribution, respectively. 
PT+ICM has the lowest TTS for problems with bimodal and
Gaussian disorder and has a clear scaling advantage. In problems with
bimodally distributed couplings, although SA results in a lower TTS for
small-sized problems, the DA and SA demonstrate similar TTSs as the
problem size increases. The performance of both SA and the DA decreases
when solving harder problem instances with Gaussian disorder, with
significantly reduced degeneracy of the ground states. However, in this
case, the DA outperforms SA even with its current precision limit.

\begin{figure*}[!htbp]
\begin{subfigure}[b]{0.49\linewidth}
\centering  
\begin{overpic}[width=1\linewidth]{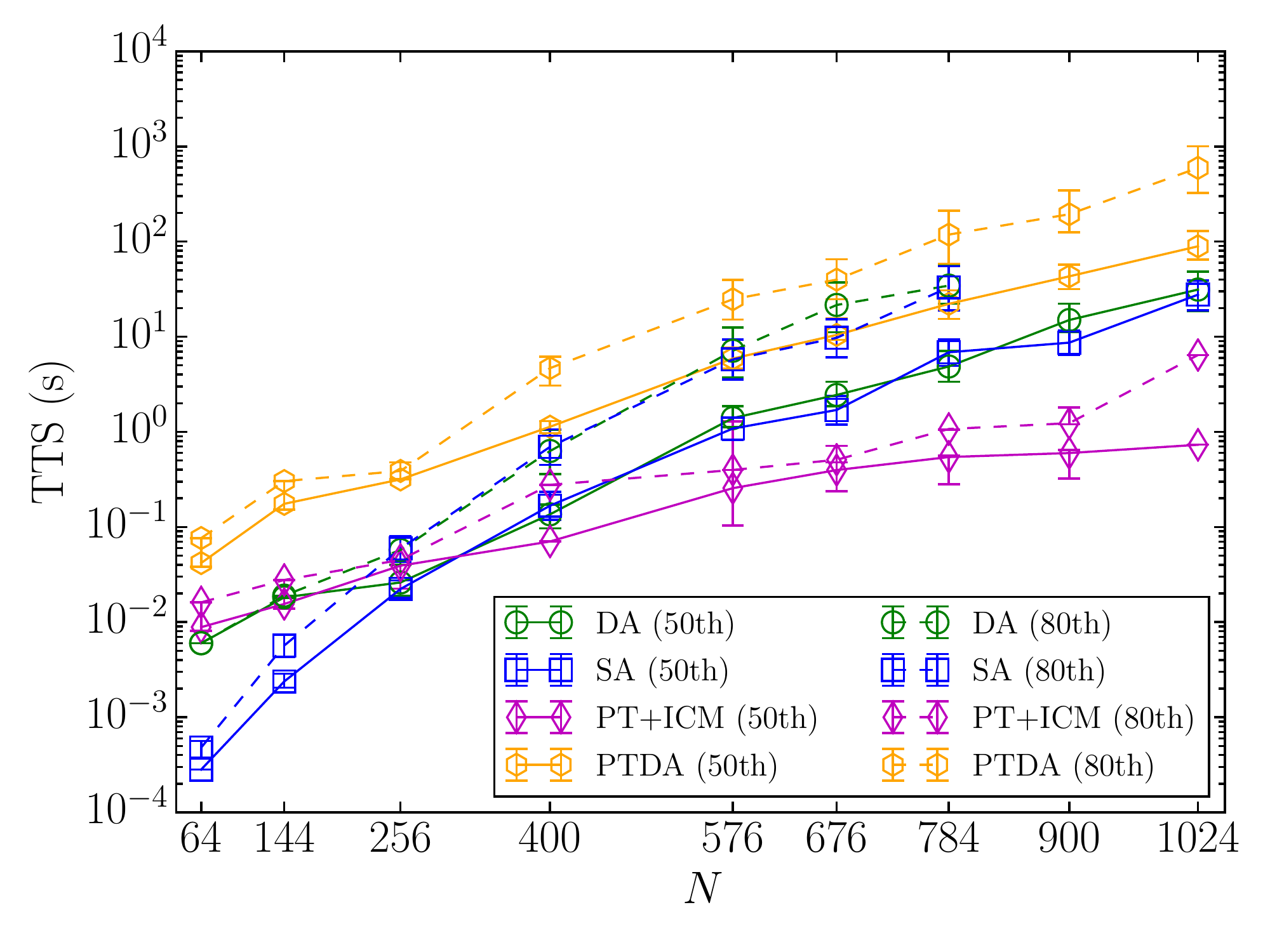} 
\put(82,66){{(a)}}
\end{overpic}
\caption{}
\label{fig:tts_2d_bimodal}
\end{subfigure} 
\begin{subfigure}[b]{0.49\linewidth}
\centering
\begin{overpic}[width=1\linewidth]{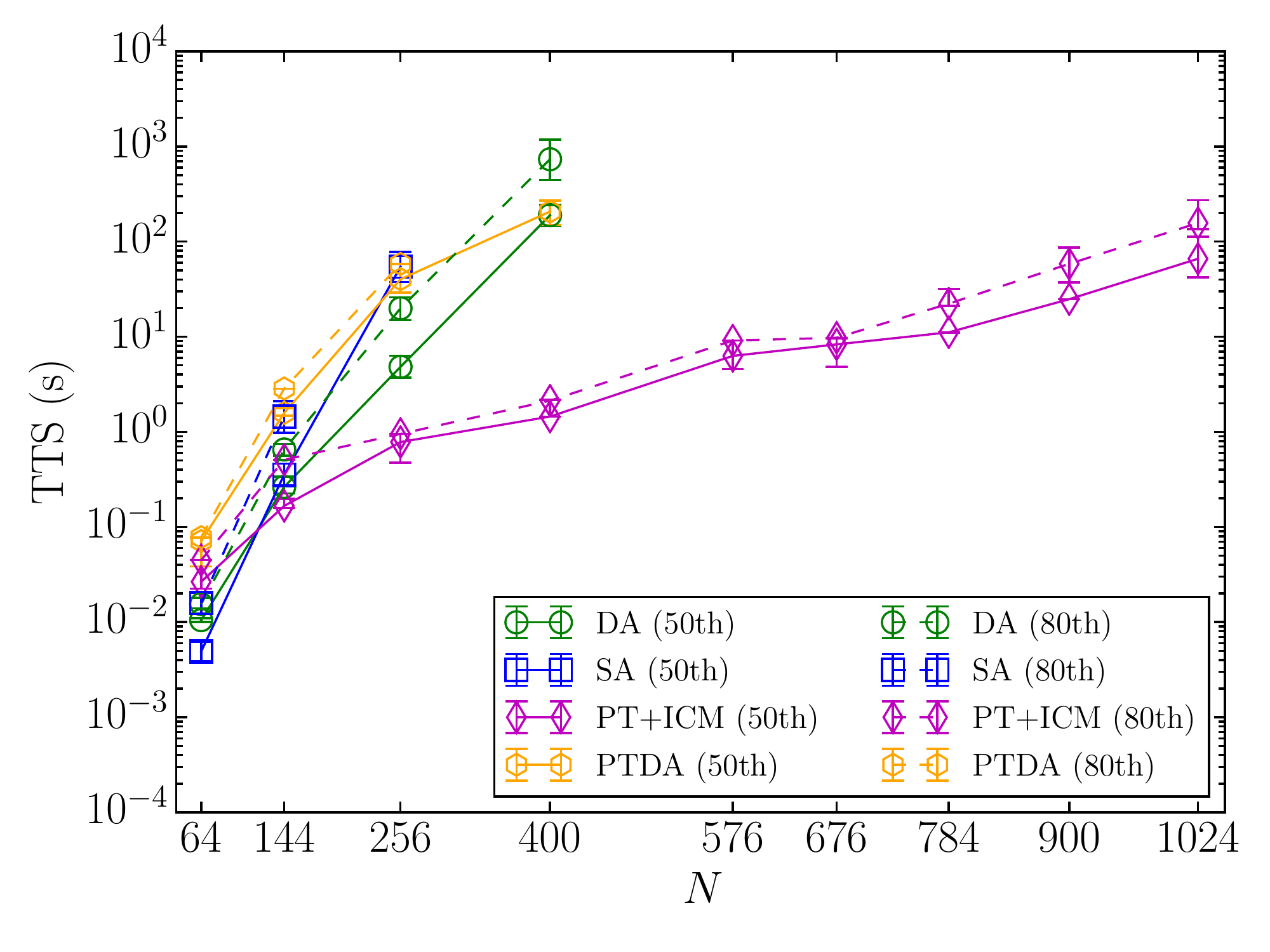} 
\put(82,66){{(b)}}
\end{overpic}
\caption{}
\label{fig:tts_2d_gaussian}
\end{subfigure}
\caption{Mean, $5$th, and $95$th percentiles of the TTS distributions
(TTS$_{50}$ and TTS$_{80}$) for {{(a)}}  2D-bimodal and {{(b)}}
2D-Gaussian spin-glass problems. For some problem sizes $N$, the
percentiles are smaller than the symbols.}
\label{fig:tts_2d}
\end{figure*}

The PTDA yields higher TTSs than the DA in both cases of bimodal and Gaussian 
disorder, likely due to the CPU overhead of performing parallel tempering moves. 
Considering the number of problems solved to optimality, the PTDA outperforms the 
DA when solving 2D spin-glass problems with bimodal couplings; however, 
as shown in Fig.~\ref{fig:tts_2d}, the PTDA solves fewer problems with Gaussian disorder. 

In order to estimate the $q$-th percentile of the TTS distribution, at least $q\%$ of problems 
should be solved to their corresponding reference energies. If there is no point for a given 
problem size and an algorithm in Fig. \ref{fig:tts_2d}, it means that enough instances have not 
been solved to their reference energies and we therefore could not estimate the TTS percentiles. 
Increasing the number of iterations to $10^7$ in the DA and the PTDA, we could not solve more than 
$80\%$ of the 2D-Gaussian problem instances with a size greater than or equal to $400$. Therefore, 
to gather enough statistics to estimate the $80$th percentile of the TTS, we have increased the number 
of iterations to $10^8$ in the DA. However, because of the excessive resources needed, we have 
not run the PTDA with such a large number of iterations. 

\subsubsection*{2D-bimodal} 

In Fig.~\ref{fig:ps_da_sa_2d_bimodal}, we observe that for a given
problem size and number of sweeps, the DA reaches higher success
probabilities than SA. As the problem size increases, the difference
between the mean success probability curves of the DA and SA becomes
less pronounced. Figure \ref{fig:correlationplot_2d_bimodal} illustrates
the success probability correlation of $100$ problem instances of size
$1024$. The DA yields higher success probabilities for $52$ problem
instances out of $100$ instances solved.

For 2D-bimodal problems, the boost in the probability of updating a
single variable due to the parallel-trial scheme is not effective enough
to decrease the TTS or to result in better scaling (Fig. \ref{fig:tts_2d_bimodal}). 
Since both the DA and SA update at most one
variable at a time, increasing the probability of updating a variable in
a problem with bimodal disorder, where there are a large number of free
variables, likely results in a new configuration without lowering the
energy value (see Sec.~\ref{sec:parallel_vs_single} for details).

\begin{figure*}[!htbp]
\begin{subfigure}[b]{0.5\linewidth}
\centering  
\begin{overpic}[width=1\linewidth]{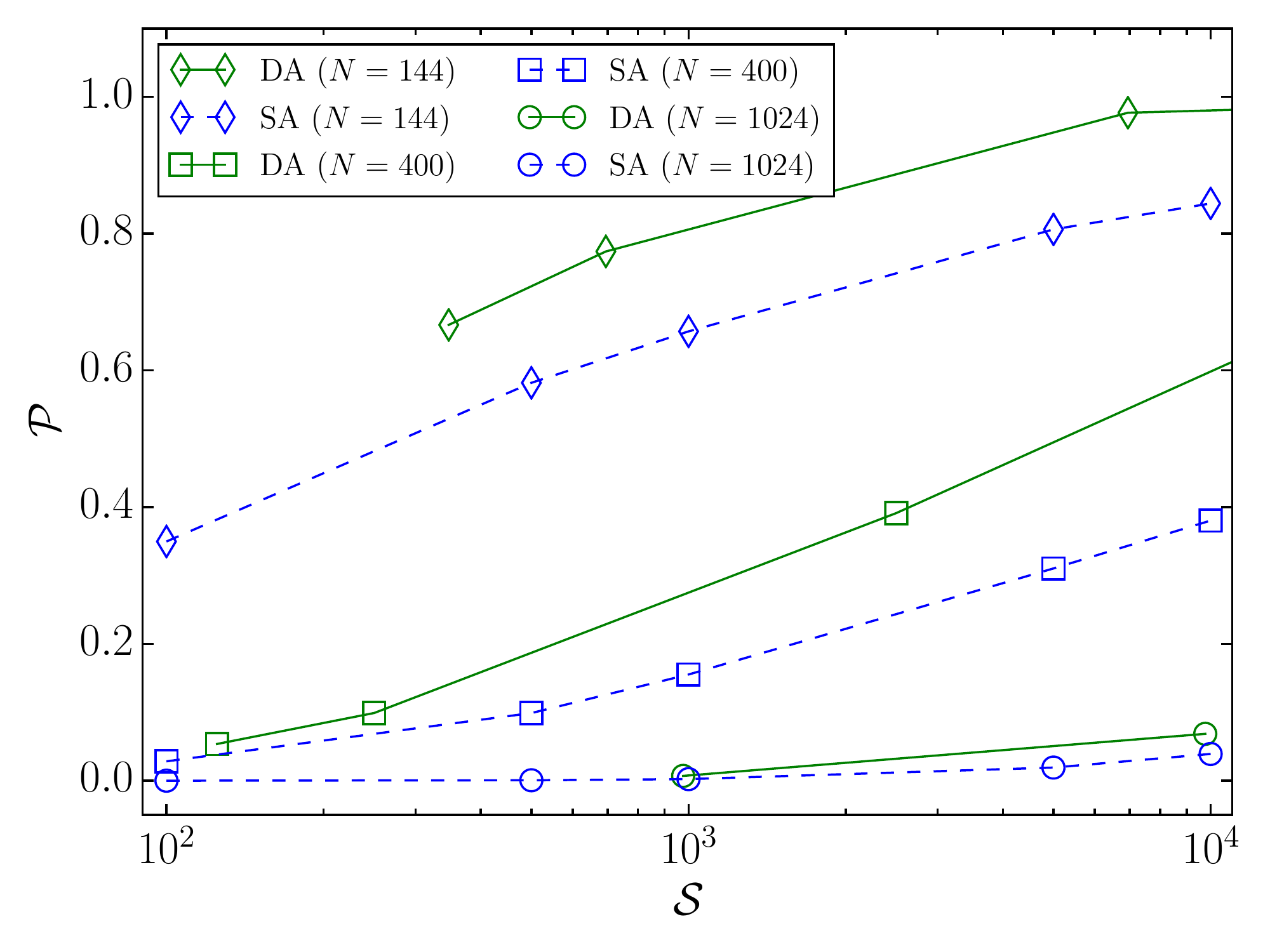} 
\put(82,67){{(a)}}
\end{overpic}
\caption{}
\label{fig:ps_da_sa_2d_bimodal}
\end{subfigure}
\begin{subfigure}[b]{0.5\linewidth}
\centering
\begin{overpic}[width=1\linewidth]{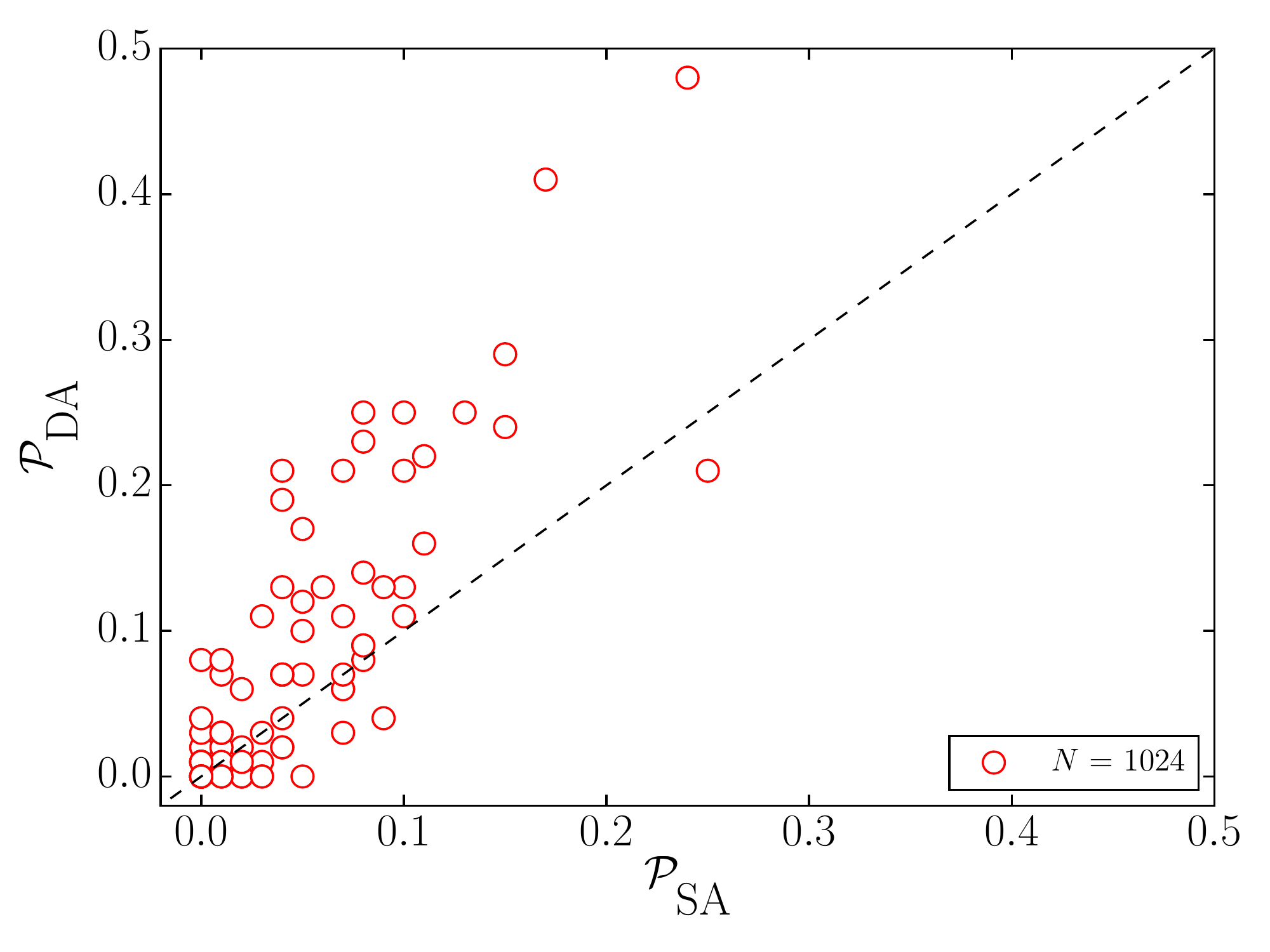} 
\put(82,67){{(b)}}
\end{overpic}
\caption{}
\label{fig:correlationplot_2d_bimodal}
\end{subfigure}
\caption{{{(a)}} Mean success probabilities ($\mathcal{P}$) of 2D-bimodal
instances vs. the number of Monte Carlo sweeps ($\mathcal{S}$) for
different problem sizes $N$ solved by the DA and SA. The error bars are
not included, for better visibility. {{(b)}} Success
probability correlation plot of $100$ 2D-bimodal instances with $N =
1024$ variables, where the number of sweeps is $10^4$ for SA and the
number of iterations is $10^7$ for the DA. Because each sweep
corresponds to $N = 1024$ iterations, the number of sweeps is $\simeq
10^4$ for the DA, as well.}
\label{fig:ps_2d_bimodal} 
\end{figure*}

\subsubsection*{2D-Gaussian} 

The performance of the DA and SA significantly degrades when
solving the problems with Gaussian disorder, which are harder; 
however, the DA demonstrates clear superiority over SA. Figure
\ref{fig:residual_2d_gaussian_1024} shows the residual energy
($\mathcal{E}$), which is the relative energy difference (in percent)
between the lowest-energy solution found and the reference-energy
solution, for the largest problem size, which has $1024$ variables. We
observe that the DA outperforms SA, as it results in a lower residual
energy for a given number of Monte Carlo sweeps ($\mathcal{S} \leq 10^5$).
Furthermore, Fig.~\ref{fig:ps_2d_gaussian} illustrates that the
parallel-trial scheme is more effective for this class of problems,
which could be due to the decrease in the degeneracy of the ground
states.  In Fig.~\ref{fig:ps_da_sa_2d_gaussianl}, we observe that the
difference between mean success probabilities, for a problem of size
$144$ with $10^4$ Monte Carlo sweeps, is larger compared to the bimodal disorder
(Fig.~\ref{fig:ps_da_sa_2d_bimodal}). The success probability
correlation of $100$ 2D-Gaussian problem instances with $400$ variables in
Fig.~\ref{fig:correlationplot_2d_gaussian} further demonstrates that
the DA reaches higher success probabilities, which results in a lower
TTS (Fig.~\ref{fig:tts_2d_gaussian}).

\begin{figure}[!htbp]
\centering
\includegraphics[width=1\linewidth]{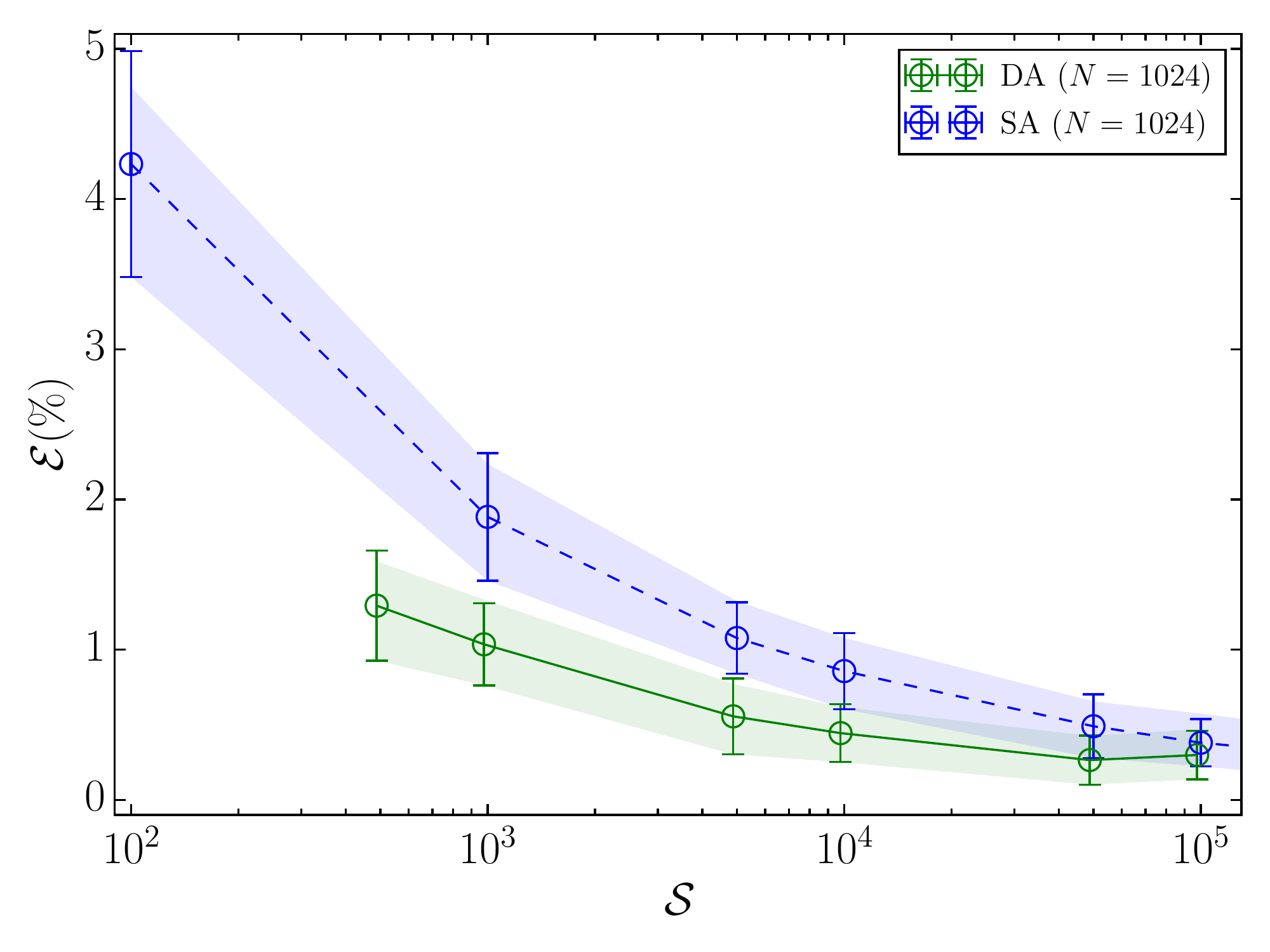} 
\caption{Mean, $5$th, and $95$th percentiles of the residual energy ($\mathcal{E}$) vs. number 
of Monte Carlo sweeps ($\mathcal{S}$) for 2D-Gaussian problem instances with $N = 1024$ 
variables solved by the DA and SA. As the number of Monte Carlo sweeps approaches infinity, 
the residual energy of both algorithms will eventually reach zero. We have run SA for up to $10^6$ 
sweeps and the DA for up to $10^8$ iterations ($10^8/1024 \sim 10^5$ sweeps). Therefore, the 
data up to $10^5$ sweeps is presented for both algorithms. We expect that by increasing the 
number of iterations to $10^9$ ($10^6$ sweeps) in the DA, the residual energy would further decrease.} 
\label{fig:residual_2d_gaussian_1024}
\end{figure}

\begin{figure*}[!htbp]
\begin{subfigure}[b]{0.49\linewidth}
\centering
\begin{overpic}[width=1\linewidth]{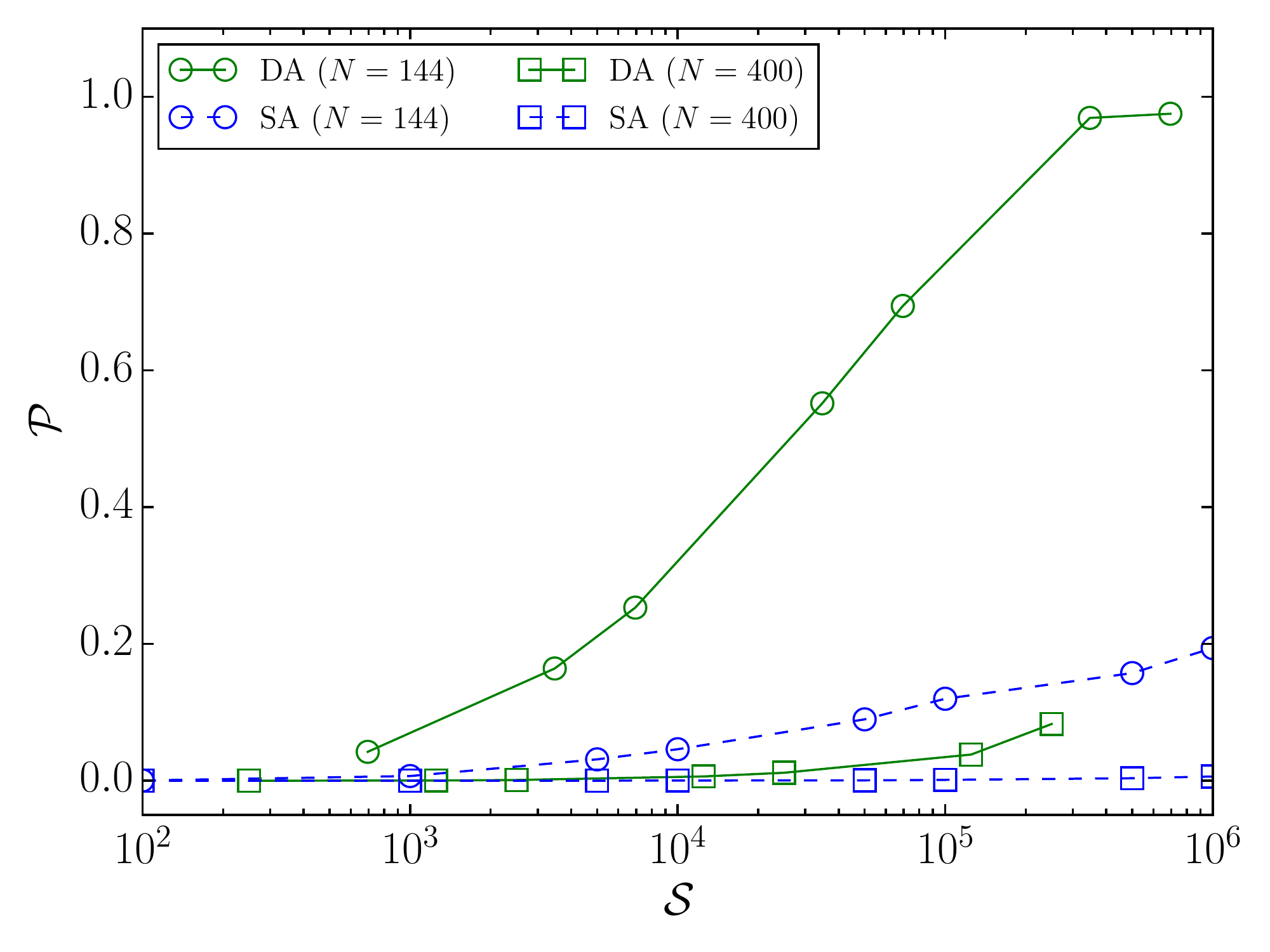} 
\put(82,68){{(a)}}
\end{overpic}
\caption{}
\label{fig:ps_da_sa_2d_gaussianl}
\end{subfigure}
\begin{subfigure}[b]{0.49\linewidth}
\centering
\begin{overpic}[width=1\linewidth]{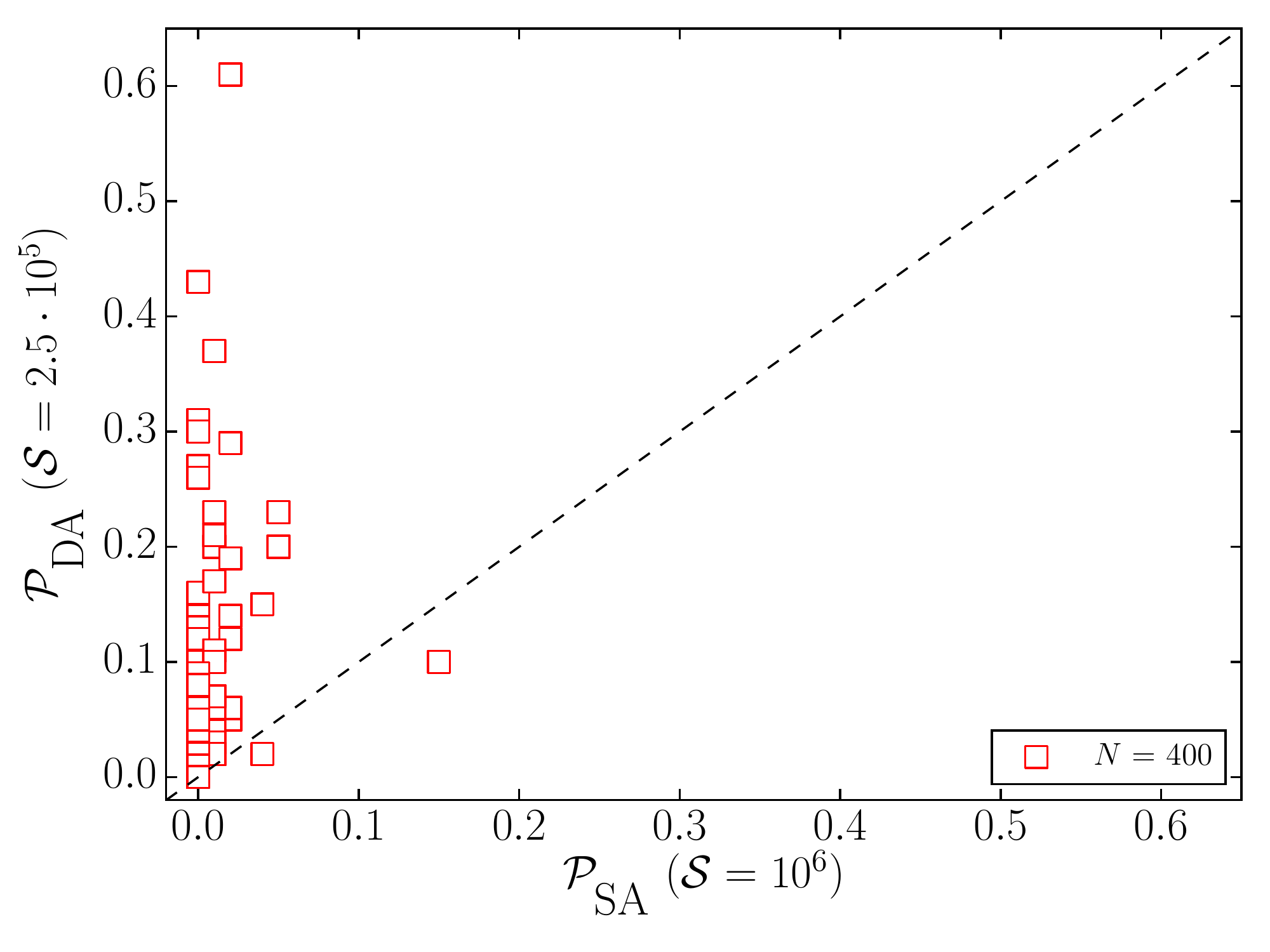} 
\put(82,68){{(b)}}
\end{overpic}
\caption{}
\label{fig:correlationplot_2d_gaussian}
\end{subfigure}
\caption{{{(a)}} Mean success probabilities ($\mathcal{P}$) vs.  the number
of Monte Carlo sweeps ($\mathcal{S}$) for 2D-Gaussian problem instances
with $N = 144$ and $N = 400$ variables solved by the DA and SA.  The
error bars are not included, for better visibility. {{(b)}} The success
probability correlation of $100$ 2D-Gaussian instances with $N = 400$
variables. The data points are obtained considering $10^6$ and $2.5
\cdot 10^5$ Monte Carlo sweeps for SA and the DA, respectively. }
\label{fig:ps_2d_gaussian} 
\end{figure*}

\subsection{SK Spin-Glass Problems}
\label{subsec:sk_results}

We have solved the SK problem instances with the DA, SA, PT, and the PTDA.
As explained in Sec.~\ref{sec:algorithms}, for the fully connected
problems, the cluster moves have not been included in PT because the clusters
of variables span the entire system. Our initial experiments 
verified that adding ICM to PT increases the computational cost without
demonstrating any scaling benefit for this problem class.

The statistics of the TTS distribution of the DA, the PTDA, and the
SA and PT algorithms are shown in Fig.~\ref{fig:tts_sk_bimodal} for
SK-bimodal problem instances. Comparing the DA to SA and PT, we 
observe that the DA yields a noticeable, consistent speedup of at least two orders 
of magnitude as we approach the largest problem size. In the fully connected problems, 
accepting a move and updating the effective local fields in a CPU implementation of a Monte Carlo
algorithm is computationally more expensive than for sparse problems.

\begin{figure*}[!htbp]
\begin{subfigure}[b]{0.5\textwidth}
\centering  
\begin{overpic}[width=0.95\textwidth]{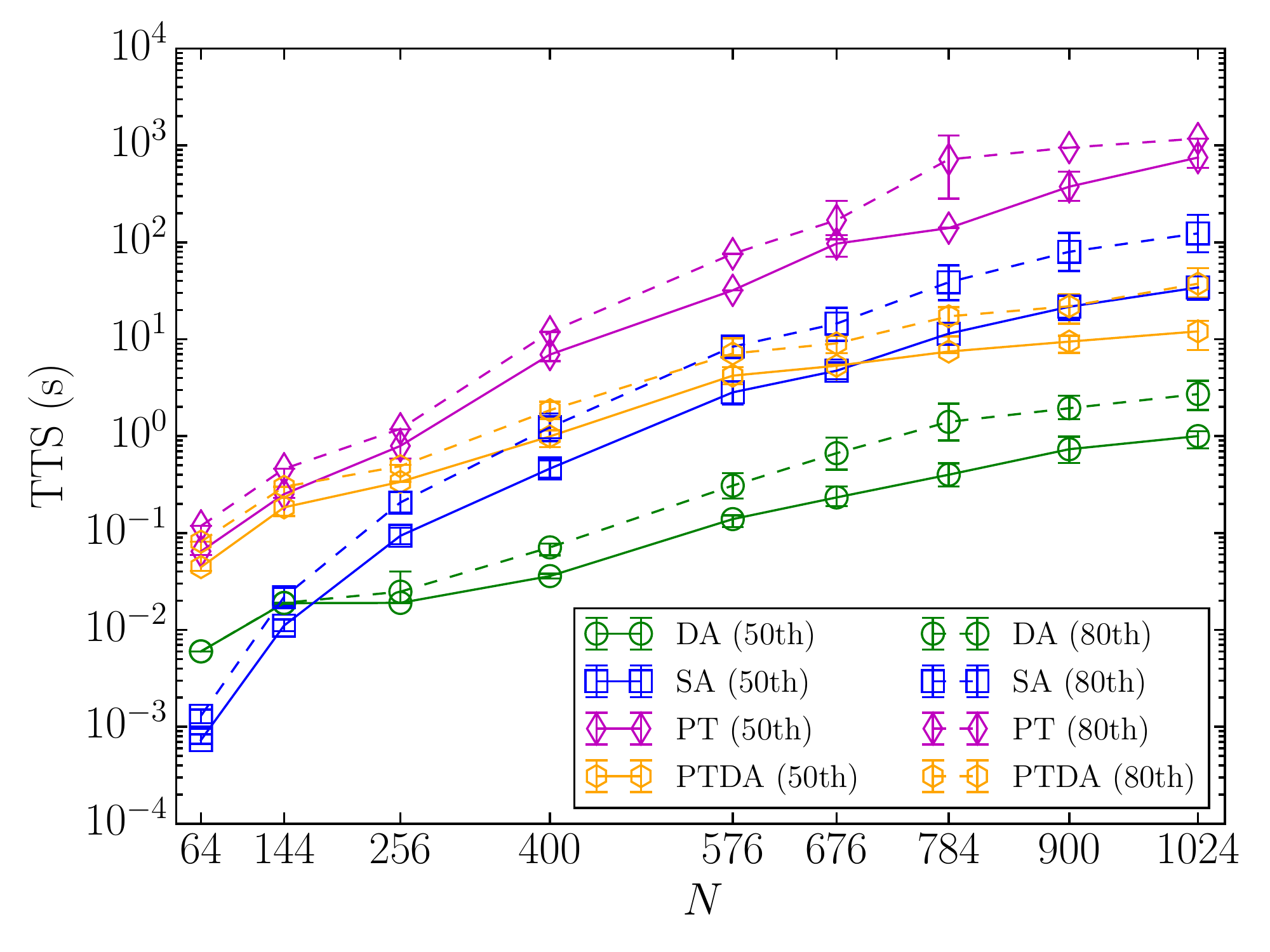} 
\put(88,67){{(a)}}
\end{overpic}
\caption{}
\label{fig:tts_sk_bimodal}
\end{subfigure}
\begin{subfigure}[b]{0.5\textwidth}
\centering
\begin{overpic}[width=1\textwidth]{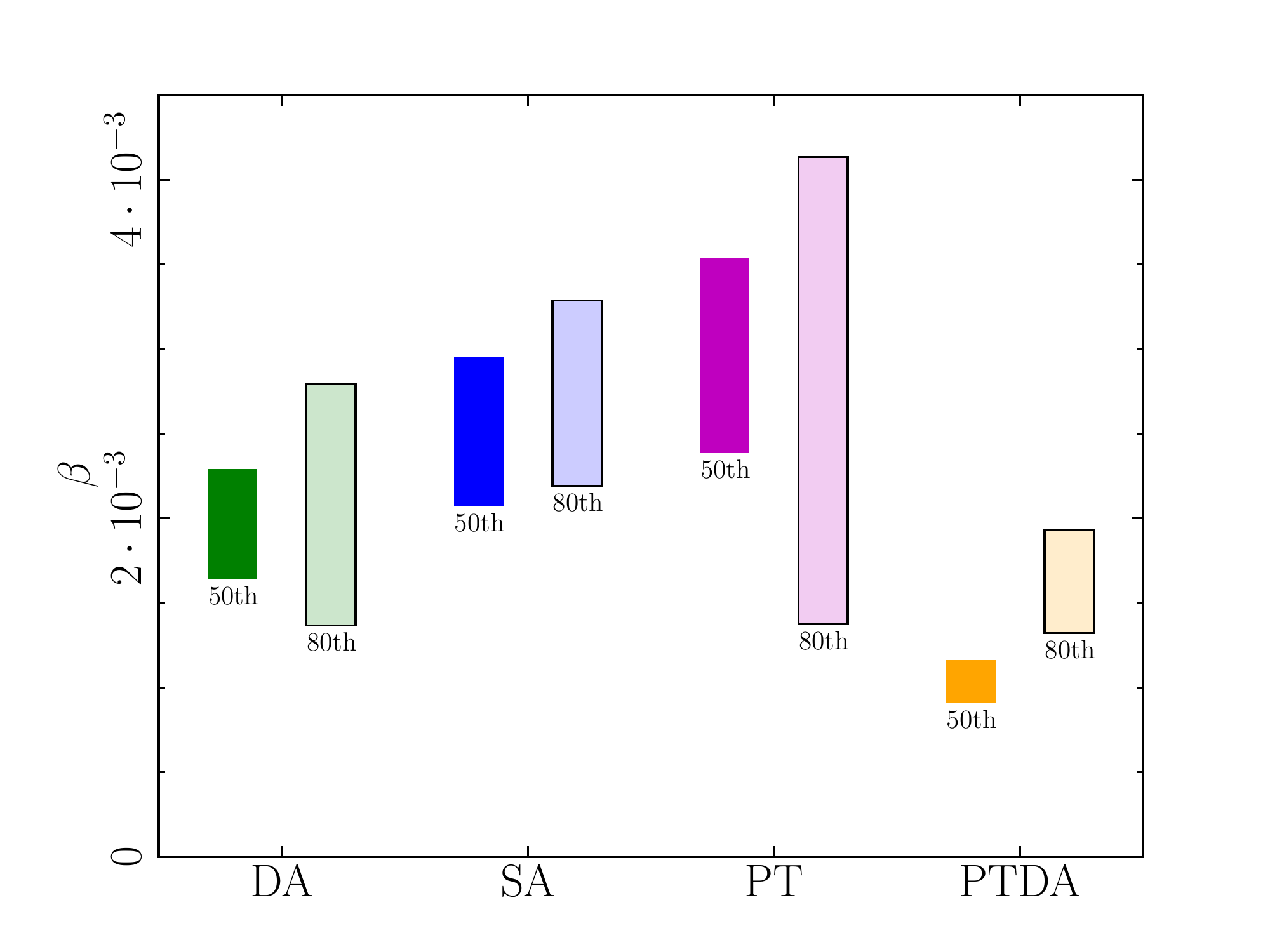} 
\put(83,64){{(b)}}
\end{overpic}
\caption{}
\label{fig:slope_sk_bimodal}
\end{subfigure}
\caption{{{(a)}} Mean, $5$th, and $95$th percentiles of the TTS
distributions (TTS$_{50}$ and TTS$_{80}$) for fully connected spin-glass
problems with bimodal disorder. For some problem sizes, the percentiles
are smaller than the symbols. {{(b)}} The $90$\% confidence interval of
the estimated scaling exponent $\beta$ based on the mean of the TTS$_{50}$
and TTS$_{80}$ for the SK-bimodal problem class.  The label below 
each rectangle represents the TTS percentile on which the confidence interval
is based. For each algorithm, the five largest problem sizes have been used to estimate the
scaling exponent by fitting a linear model to $\log_{10}\mbox{TTS}$. The $R^2$ of the fitted 
model is greater than or slightly lower than $90\%$ for each algorithm.}
\end{figure*}

Figure \ref{fig:tts_sk_bimodal} shows that each algorithm
has solved at least $80\%$ of the SK instances for all problem sizes. We
attribute this to the fact that the reference energy for the complete
graph problems is an \textit{upper bound} on the exact optimal solution.
We do not know how tight the upper bound is, but it represents, to the
best of our knowledge, the best known solution.

To obtain insights on scaling, for each algorithm, we have fit an exponential
function of the form $y = 10^{\alpha + \beta N}$, where $y$ and $N$ are
the means of the TTS distribution and the number of variables,
respectively. Figure \ref{fig:slope_sk_bimodal} shows the $90\%$
confidence interval of the estimated scaling exponent $\beta$ for the
algorithms based on the statistics of the $50$th and the $80$th
percentiles of the TTS distribution. For the $50$th percentile, we
observe that the PTDA yields superior scaling over the other three
algorithms for the problem class with bimodal disorder. For the $80$th
percentile, there is not enough evidence to draw a strong conclusion on
which algorithm scales better because the $90$\% confidence intervals of
the estimated scaling exponents overlap. However, the PTDA has the
lowest point estimate.

For the DA, SA, and PT algorithms, we have searched over a 
large number of parameter combinations to experimentally determine a
good set of parameters (see \ref{sec:appendix_parameters})
while the parameters of the PTDA have been determined automatically by the
hardware. We have further experimentally determined the optimal number
of sweeps for all four algorithms. However, we do not rule out the
possibility that the scaling of the algorithms might be suboptimal due
to a non-optimal tuning of parameters.  For example, the scaling of the
PTDA might improve after tuning its parameters and PT might exhibit
better scaling using a more optimized temperature schedule.

Figure \ref{fig:tts_sk_gaussain} illustrates the TTS statistics and the
confidence interval of the scaling exponent for SK-Gaussian problem
instances solved by the DA, SA, PT, and the PTDA. We observe that the DA 
continues to exhibit a constant speedup of at least two orders of magnitude over 
the other algorithms, with no strong scaling advantage, in solving spin-glass problems 
with Gaussian disorder.

\begin{figure*}[!htbp]
\begin{subfigure}[b]{0.5\textwidth}
\centering  
\begin{overpic}[width=0.95\textwidth]{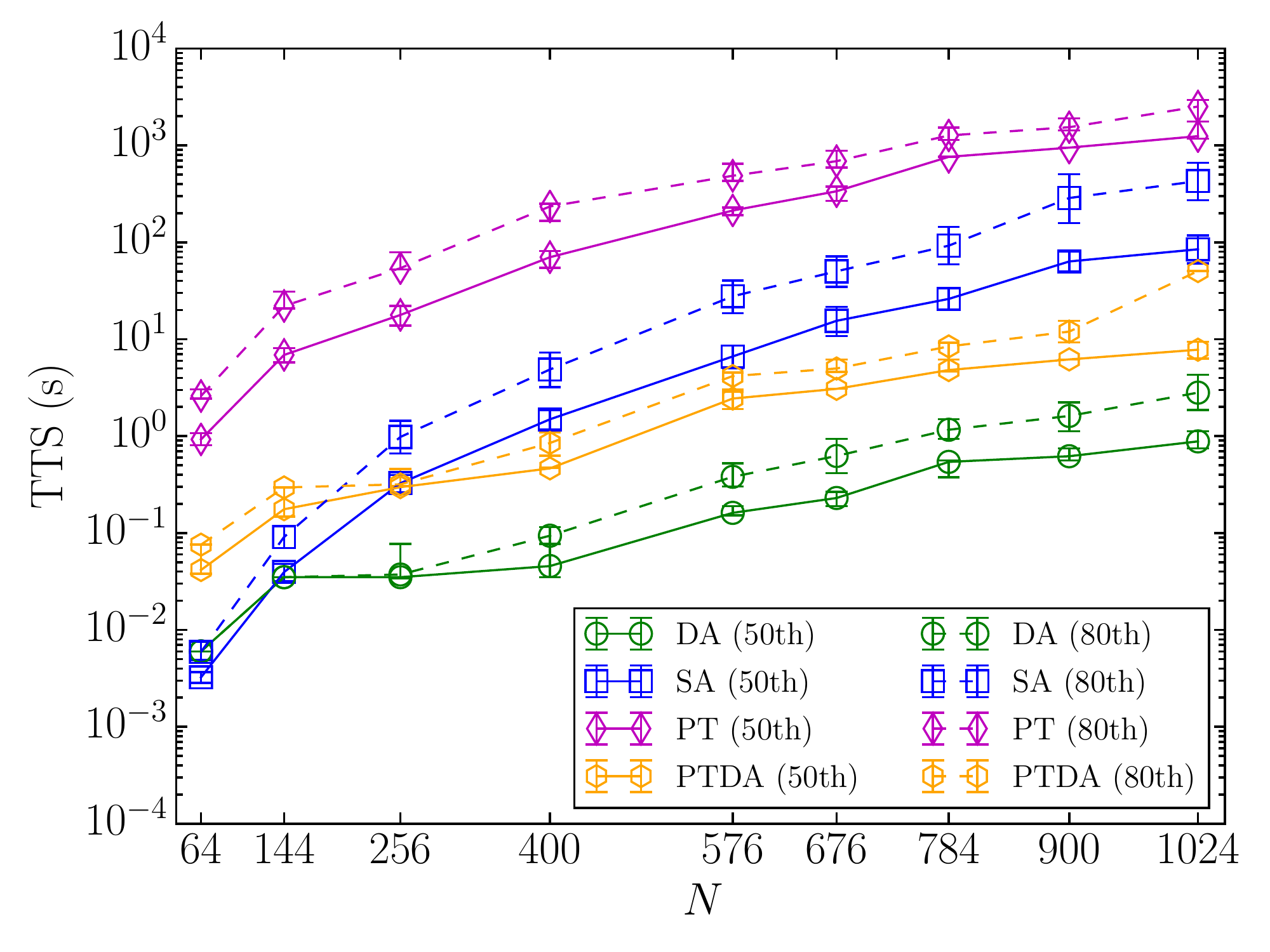} 
\put(87,68){{(a)}}
\end{overpic}
\caption{}
\end{subfigure}
\begin{subfigure}[b]{0.5\textwidth}
\centering
\begin{overpic}[width=1\textwidth]{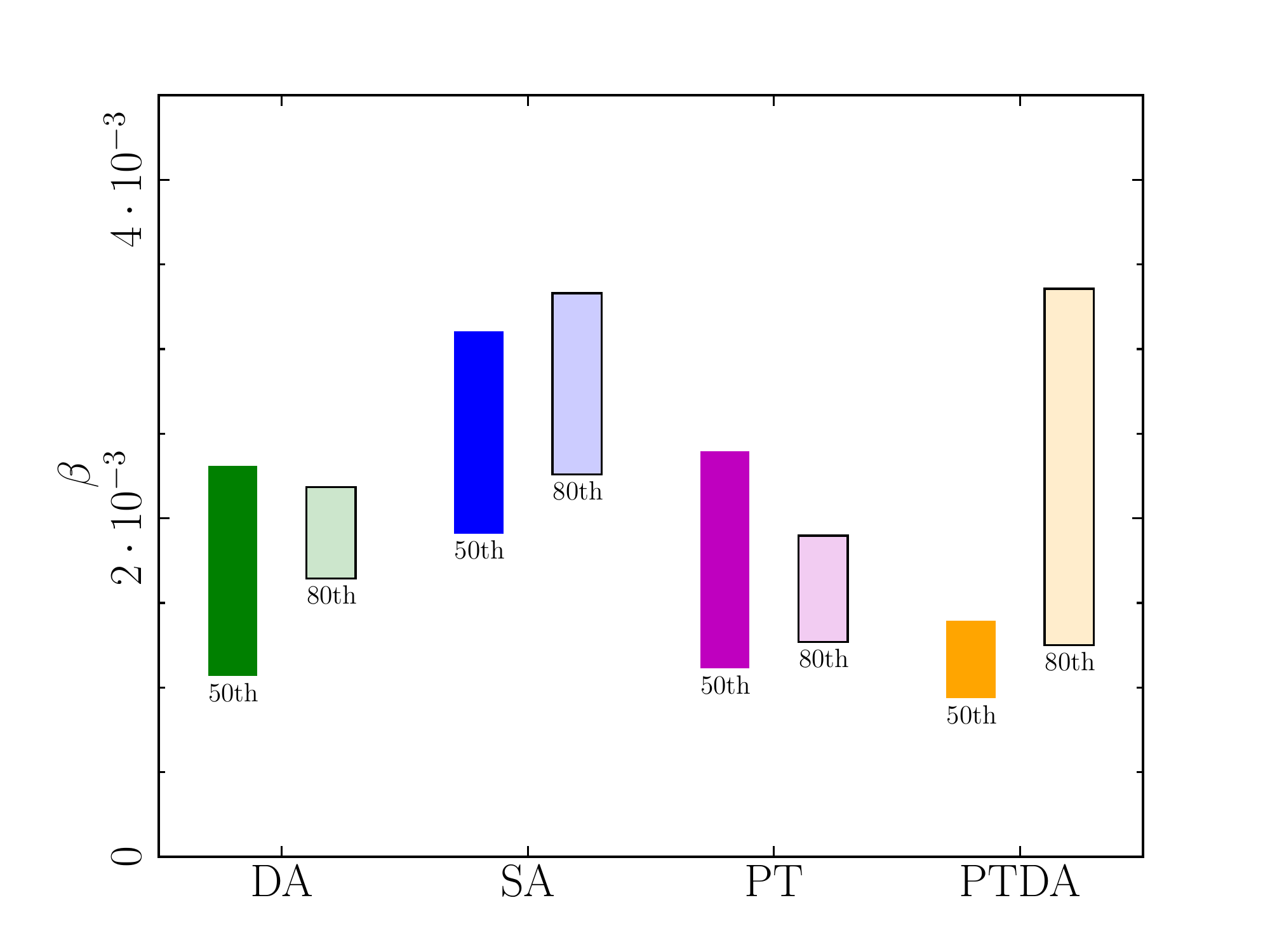} 
\put(81,64){{(b)}}
\end{overpic}
\caption{}
\label{fig:slope_sk_gaussian}
\end{subfigure}
\caption{{{(a)}} Mean, $5$th, and $95$th percentiles of the TTS
distributions (TTS$_{50}$ and TTS$_{80}$) for fully connected spin-glass
problems with Gaussian disorder. For some problem sizes, the percentiles
are smaller than the symbols. {{(b)}} The $90$\% confidence interval of
the estimated scaling exponent $\beta$ based on the mean of the TTS$_{50}$
and TTS$_{80}$ for the SK-Gaussian problem class. The label below
each rectangle represents the TTS percentile on which
the confidence interval is based. For each algorithm, the five largest problem sizes 
have been used to estimate the scaling exponent by fitting a linear model to $\log_{10}\mbox{TTS}$. 
The $R^2$ of the fitted model is greater than or slightly lower than $90\%$ for each algorithm.}
\label{fig:tts_sk_gaussain}
\end{figure*}

\subsubsection*{The DA versus SA} 

The DA results in lower TTSs than SA for both SK-bimodal and
SK-Gaussian problem instances. The reasons for this behaviour are
twofold. First, the anneal time for the DA is independent of the number
of variables and the density of the problem, whereas the computation
time of a sweep in SA increases with the problem size and the problem
density.  Second, as shown in Fig.~\ref{fig:sk_sa_da_correlation_plot},
the parallel-trial scheme significantly improves the success probability
in fully connected spin-glass problems of size $1024$ with both bimodal
and Gaussian disorder. As expected, the boost in the low-degeneracy
problem instances (with Gaussian coefficients) is higher.

\begin{figure}[!h]
\centering  
\includegraphics[width=1\linewidth]{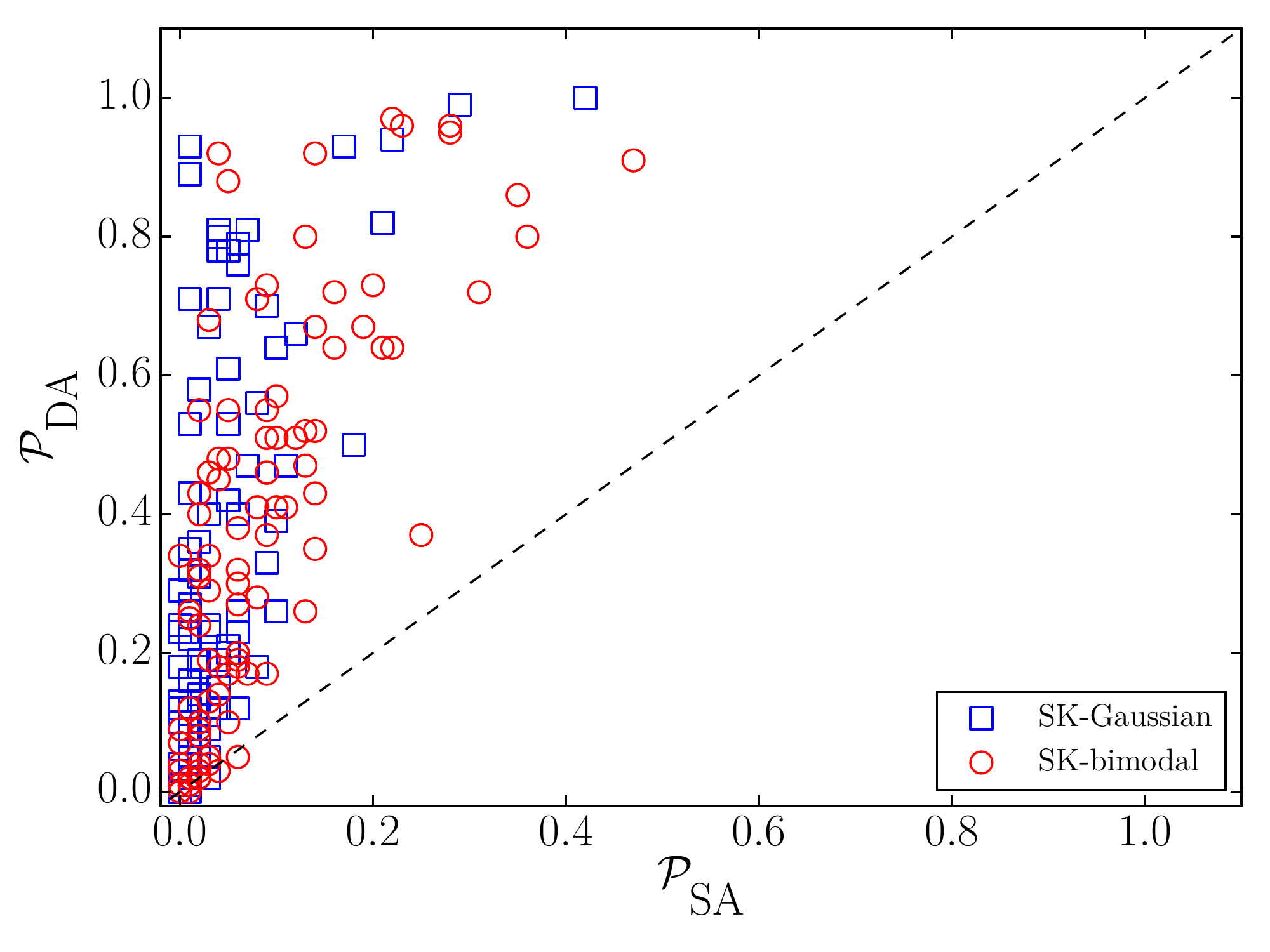} 
\caption{Success probability correlation ($\mathcal{P}$) of $100$ SK
instances of size $1024$ with bimodal and Gaussian disorder. The number
of Monte Carlo sweeps in SA is $10^4$ and the number of iterations in
the DA is $10^7$, corresponding to $\simeq 10^4$ Monte Carlo sweeps.}
\label{fig:sk_sa_da_correlation_plot} 
\end{figure}

Although the confidence intervals of the scaling exponents overlap,
considering the statistics of the TTS$_{80}$, the DA yields lower point
estimates than SA for SK-bimodal and SK-Gaussian problems. In
particular, $\beta = 0.0021(7)$ [$0.0019(3)$] for the DA with bimodal
[Gaussian] disorder, whereas $\beta = 0.0027(6)$ [$0.0028(5)$] for SA
with bimodal [Gaussian] disorder, thus providing a weak scaling
advantage.

Our results on spin-glass problems with Gaussian disorder further indicate that the $16$-bit 
precision of the hardware used in this study is not a limiting factor because the DA outperforms 
SA on instances of these problems. Since there is a high variance in the couplers of spin-glass 
instances with Gaussian disorder, we expect that the energy gap between the ground state and 
the first-excited state is likely greater than $10^{-5}$ and, as a result, the scaling/rounding 
effect is not significant \cite{precision}. Our experimentation data on the prototype of the second-generation Digital Annealer, 
which has $64$ bits of precision on both biases and couplers, also confirms that the higher precision by itself 
does not have a significant impact on the results presented in this paper. We leave a presentation 
of our experimental results using the second-generation hardware for  future work.

\subsection{Spin-Glass Problems with Different Densities}
\label{subsec:various_density}

Our results for the two limits of the problem-density spectrum suggest that the DA 
exhibits similar TTSs to SA on sparse problems, and outperforms SA on fully connected problems 
by a TTS speedup of approximately two orders of magnitude. To obtain a deeper understanding of the relation
between the performance and the density, we have performed an 
experiment using random problem graphs with nine different densities. For each problem density, 
$100$ problem instances with $1024$ variables have been generated based on the Erd\H os--R\'enyi model \cite{erdos59}, with 
bimodally distributed coupling coefficients, and zero biases. The parameters of the 
DA and SA have been experimentally tuned for each of the nine problem densities 
(see \ref{sec:appendix_parameters}). 

Figure~\ref{fig:tts_varying_density} shows the statistics of the TTS of the DA and SA for different 
problem densities. The TTS results for 2D-bimodal ($d = 0.4 \%$) and SK-bimodal ($d = 100 \%$) 
for a problem size of $1024$, representing the limits of the density spectrum, are also included. 
The DA has lower TTSs than SA for all problem densities except for the sparsest problem set---2D-bimodal. 
Not enough 2D-bimodal instances were solved to optimality using the DA and SA in order to estimate 
the statistics of the TTS$_{80}$ distributions. 

\begin{figure}[!htbp]
\centering  
\includegraphics[width=1\linewidth]{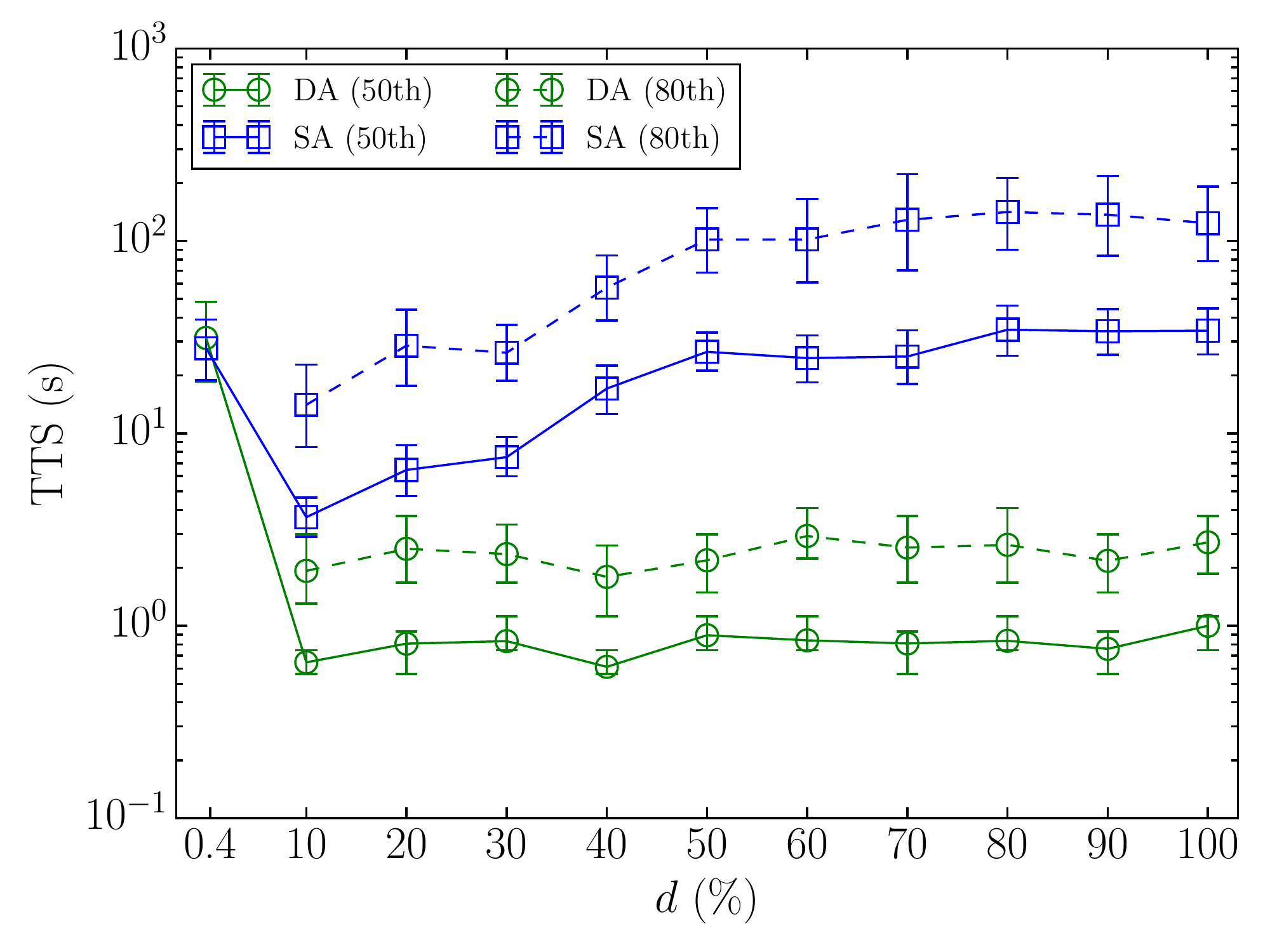} 
\caption{Mean, $5$th, and $95$th percentiles of the TTS
distributions (TTS$_{50}$ and TTS$_{80}$) for spin-glass problems 
with different densities ($d$).}
\label{fig:tts_varying_density}
\end{figure}

Figure~\ref{fig:varying_density_plot} shows the success 
probabilities of $100$ spin-glass problem instances of size $1024$ with 
different densities solved by the DA and SA. The DA has higher success probabilities than SA
by a statistically significant margin for all of the densities except for the sparsest problem set---2D-bimodal. 
We interpret these results as being due to both the increase in the success
probabilities from using a parallel-trial scheme and the constant time 
required to perform each Monte Carlo step on the DA hardware architecture. 

\begin{figure}[!htbp]
\centering  
\includegraphics[width=1\linewidth]{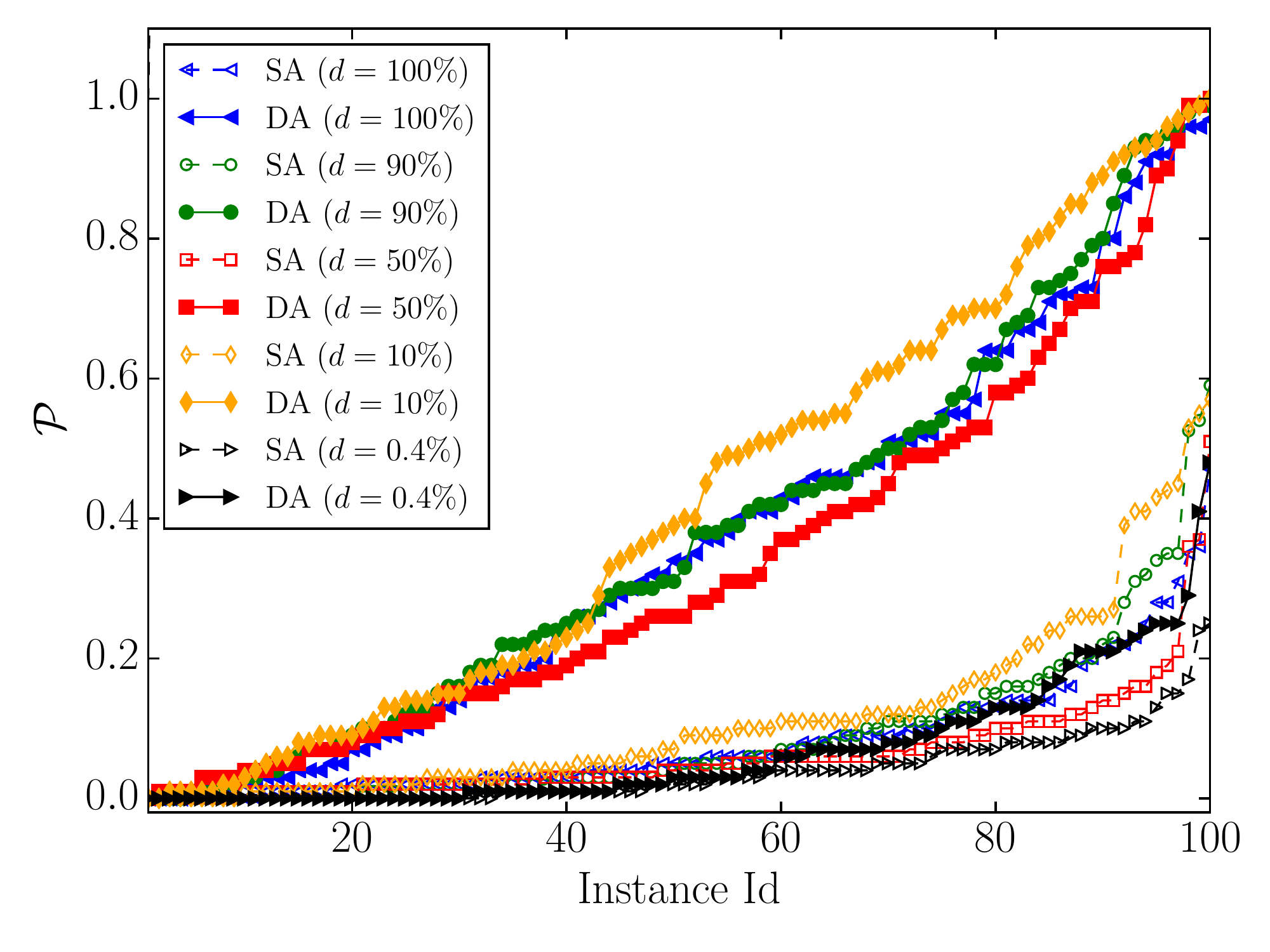} 
\caption{Success probabilities ($\mathcal{P}$) 
of $100$ spin-glass problem instances with different densities solved by the SA and the DA algorithms. 
The number of Monte Carlo sweeps in SA is $10^4$ and the number of iterations in the DA is $10^7$, 
corresponding to $\simeq 10^4$ Monte Carlo sweeps.}
\label{fig:varying_density_plot}
\end{figure}

\section{Conclusions and Outlook}
\label{sec:conclusions}

In this work we have compared the performance of the Digital Annealer
(DA) and the Parallel Tempering Digital Annealer (PTDA) to parallel
tempering Monte Carlo with and without isoenergetic cluster moves
(PT+ICM and PT, respectively) and simulated annealing (SA) using random
instances of sparse and fully connected spin-glass problems, with
bimodal and Gaussian disorder.

Our results demonstrate that the DA is approximately two orders of
magnitude faster than SA and PT in solving dense problems,
while it does not exhibit a speedup for sparse problems. In the latter
problem class, the addition of cluster updates to the PT
algorithm is very effective in traversing the energy barriers,
outperforming algorithms that act on a single flip neighbourhood, such
as the DA and SA. For dense problems, the efficiency of the cluster
moves diminishes such that the DA is faster, due to the parallel-trial
scheme combined with the massive parallelization that is possible on
application-specific CMOS hardware. Our results further support the
position that the DA has an advantage over SA on random spin-glass 
problems with densities of $10\%$ or higher. 

In Sec.~\ref{sec:parallel_vs_single} we demonstrate that parallel-trial
Monte Carlo can offer a significant boost to the acceptance
probabilities over standard updating schemes.  Furthermore, we show that
this boost vanishes at high temperatures and is diminished for problems
with high ground-state degeneracy. Our benchmarking results further
support the view that the parallel-trial scheme is more effective in
solving problems with low ground-state degeneracy because an accepted
move is more likely not only to change the state configuration, but also
to lower the energy value.

In the current early implementation of the PTDA, the TTS is higher than
it is likely to be in the future, due to the CPU overhead in performing
PT moves. However, the PTDA algorithm demonstrates better scaling than
the other three algorithms for a fully connected spin-glass problem of
average computational difficulty, with bimodal couplings.

In the next generation of the Digital Annealer, the hardware
architecture is expected to allow the optimization of problems using
up to 8192 fully connected variables. In addition, the annealing time
is expected to decrease, and we conjecture that the TTS might decrease
accordingly. Finally, we expect the replica-exchange moves in the PTDA
to be performed on the hardware, which could improve the performance of
the PTDA.

Our results demonstrate that pairing application-specific CMOS hardware 
with physics-inspired optimization methods results in extremely efficient,
special-purpose optimization machines. Because of their fully connected
topology and high digital precision, these machines have
the potential to outperform current analog quantum optimization
machines. Pairing such application-specific CMOS hardware with a fast
interconnect could result in large-scale transformative optimization
devices.  We thus expect future generations of the Digital Annealer to
open avenues for the study of fundamental physics problems and industrial
applications that were previously inaccessible with conventional CPU
hardware.

\section*{Conflict of Interest Statement} 
Authors MA, GR, and EV are employed by the company 1QBit. Authors TM and HT are 
employed by the company Fujitsu Laboratories Ltd. Author HGK was employed by 1QBit 
when this research was completed and is currently employed by the company Microsoft Research. 

The authors declare that there is no further commercial or financial relationship that could be construed 
as a potential conflict of interest.

\section*{Author Contributions}
Authors MA, GR, and HGK developed the methodology, implemented the code, performed the 
experiments, analyzed the results, and wrote the manuscript. Author EV partially contributed to 
implementing the code and conducting the experiments. Authors TM and HT carried out the 
experiments related to the PTDA algorithm. 

\section*{Funding}
This research was supported by 1QBit, Fujitsu Laboratories Ltd., and
Fujitsu Ltd. The research of HGK was supported by the
National Science Foundation (Grant No.~DMR-1151387). HGK's research
is based upon work supported by the Office of the Director of National
Intelligence (ODNI), Intelligence Advanced Research Projects Activity
(IARPA), via Interagency Umbrella Agreement No.~IA1-1198. The views and
conclusions contained herein are those of the authors and should not be
interpreted as necessarily representing the official policies or
endorsements, either expressed or implied, of the ODNI, IARPA, or the
U.S.~Government.  The U.S.~Government is authorized to reproduce and
distribute reprints for Governmental purposes notwithstanding any
copyright annotation thereon.

\section*{Acknowledgement} 
The authors would like to thank Salvatore Mandr\`a for
helpful discussions, Lester Szeto, Brad Woods, Rudi Plesch, Shawn Wowk,
and Ian Seale for software development and technical support, Marko
Bucyk for editorial help, and Clemens Adolphs for reviewing the manuscript. 
We thank Zheng Zhu for providing us with his implementation of 
the PT+ICM algorithm \cite{zhu:15b,zhu:16y}, and
Sergei Isakov for the use of his implementation of the SA algorithm
\cite{isakov:15}. HGK would like to thank Bastani
Sonnati for inspiration.

 \appendix
\section{Simulation Parameters}
\label{sec:appendix_parameters}
In this section we present the parameters of the algorithms used to solve 
two-dimensional and SK spin-glass problems with bimodal and Gaussian disorder, then explain the parameters used to calculate the TTSs of the DA and SA when solving spin-glass problems with different densities.

\subsection*{2D \& SK Spin-Glass Problems}
\label{subsec:appendix_2d_sk}

For the DA, SA, and PT (PT+ICM) and each problem class, 
we have performed a grid search in the parameter space to
determine the best parameters, using
a subset of problem instances. The subset of instances used for
parameter tuning include instances of size $576$, $784$, and $1024$.
The number of instances solved to optimality, the success probability,
and the residual energy have been used to select the best parameter
combination for the benchmarking study.

In order to find the optimal TTS for a given problem size, we have varied the
number of sweeps (iterations in the DA and the PTDA) and have used the
procedure shown in Algorithm \ref{TTS} to find the empirical
distribution of the TTS for each number of sweeps. We have then found the number
of sweeps for which the TTS distribution has the lowest mean and report
the statistics of that distribution as the optimal TTS results. In the
calculation of the TTS, we have excluded the initialization and
post-processing times. The time spent on replica-exchange moves,
currently performed via CPU, has been considered to be part of the PTDA's
execution time. The time that it takes to execute one run of SA and PT
(and PT+ICM) has been measured using \textit{r4.8xlarge} Amazon EC2
instances, which consist of Intel Xeon E5-2686 v4 (Broadwell)
processors.

To set the grid for the high and the low temperatures, we have simulated the
distribution of the energy differences associated with proposed moves
and search in the vicinity of the $5$th ($80$th) to the $10$th ($85$th)
percentiles of this distribution to find the best-performing low (high)
temperature value. In PT (PT+ICM), because we are able to measure
different quantities during the simulation, we have further ensured that the
highest temperature has been chosen such that the Monte Carlo acceptance
probabilities are between $0.6$ and $0.8$. The parameter values used in
each algorithm are outlined below.

\subsubsection*{The DA parameters} 

For all experiments, we have used $100$ runs, each starting at a vector of
zeros. The temperature schedule is linear in the inverse temperature,
and the temperature has been adjusted after every iteration. The DA uses the
Metropolis criterion to accept Monte Carlo moves. It is worth mentioning
that our early experimentation has suggested better performance for
the linear inverse temperature schedule than the exponential temperature
schedule.

Our investigation of different parameter combinations has further shown
that the performance of the DA on spin-glass problems is indifferent to
the dynamic offset mechanism. Therefore, we turn this feature off for
our final experimentation. The high (\Th) and the low (\Tl) temperatures
used for each problem class are shown in Table
\ref{table:final_parameters}.

\begin{table}[h]  
\centering
\footnotesize
\begin{tabular*}{\columnwidth}{@{\extracolsep{\fill}} c r r r r r r r r }
\hline
\hline
\multirow{2}{*}{} 
					 & \multicolumn{2}{c}{2D-bimodal} & \multicolumn{2}{c}{2D-Gaussian} & \multicolumn{2}{c}{SK-bimodal} & \multicolumn{2}{c}{SK-Gaussian} \\
					 & \multicolumn{1}{c}{\Th} 	& \multicolumn{1}{c}{\Tl} 	& \multicolumn{1}{c}{\Th} 	& \multicolumn{1}{c}{\Tl} 	&  \multicolumn{1}{c}{\Th}	& \multicolumn{1}{c}{\Tl} 	& \multicolumn{1}{c}{\Th} 	& \multicolumn{1}{c}{\Tl} \\ \hline
DA 					& 	2	&  0.66	& 	$2 \cdot 10^4$	         & $4 \cdot 10^2$		& 	40	&	4	& 	$2 \cdot 10^5$	& 	$10^4$	\\ 
SA 					& 	10	&  0.33	& 	$10^6$	 & 	$10^4$	& $\sqrt{N}$ &	1	& 	$10^7$	& 	$10^5$	\\ 
PT  				& 	2	&  0.33	& 	$10^6$	         & $2 \cdot 10^4$		& 	80	&	2	& 	$10^7$	& 	$5 \cdot 10^5$	\\ 
\hline
\hline
 \end{tabular*}
 \caption{
High (\Th) and low (\Tl) temperatures used in the DA, SA, and PT
runs. Note that the row marked with ``PT'' also includes the
PT+ICM parameters. The low and the high temperatures of SA for
2D-bimodal and SK-bimodal are selected according to
Refs.~\cite{isakov:15,venturelli:15a}. The temperature values used in 
our simulations are unitless.}
\label{table:final_parameters}
\end{table}

\subsubsection*{SA parameters} 

Each instance has been solved $100$ times using SA and the temperature schedule has been set to be linear in the inverse temperature, which is a typical choice
for SA in the literature \cite{isakov:15,venturelli:15a}.  The high
(\Th) and the low (\Tl) temperatures used for each problem class are
shown in Table \ref{table:final_parameters}.
 
\subsubsection*{PT (PT+ICM) parameters} 

Although the performance of replica-exchange algorithms is significantly
dependent on the choice of temperature schedule, the temperatures at each replica have been set based on the commonly used geometric schedule
\cite{katzgraber:06a}. After determining the low and the high
temperatures, the number of replicas has been chosen such that the replica-exchange acceptance probabilities are above $0.2$. The number of
replicas has been set to $25$, $60$, $50$, and $60$, respectively, for the 2D-bimodal, SK-bimodal, 2D-Gaussian, and SK-Gaussian
instances. In contrast to the runs of the DA and SA, the replicas in
PT (PT+ICM) are not independent and to calculate the TTS, the
whole PT (PT+ICM) algorithm has been repeated $30$ times for each instance.
This is time consuming, so each run (repeat) has been stopped immediately if
the reference solution has been found. The high (\Th) and the low (\Tl)
temperatures used for each problem class are shown in Table~\ref{table:final_parameters}.

\vspace{0.5cm}
\subsubsection*{The PTDA parameters} 

The number of replicas has been set to $40$ for both the SK-bimodal and
SK-Gaussian problem instances, and the dynamic offset feature has been turned
off. The high and the low temperatures and the temperature schedule are
set internally by an automatic parameter-tuning strategy (see also
Sec.~\ref{sec:algorithms}). As done for PT, since the replicas are
dependent, the whole algorithm has been repeated $30$ times to have enough
observations to calculate the TTS for each instance.

\subsection*{Spin-Glass Problems with Different Densities}
\label{subsec:appendix_diff_densities}

A grid-search approach on a subset of instances has been used to tune the parameters of the DA and SA
for spin-glass problems, as explained in the previous section, separately for each density.
The parameters of the DA and SA 
are the same as the ones used for 2D and SK spin-glass problems,
except for the temperature values that are given below.

\begin{table}[ht]  
\centering
\footnotesize
\begin{tabular*}{\columnwidth}{@{\extracolsep{\fill}} c r r r r }
\hline
\hline
\multirow{2}{*}{} 
				        & \multicolumn{2}{c}{DA} & \multicolumn{2}{c}{SA} \\
$d\,(\%)$			        & \multicolumn{1}{c}{\Th} 	& \multicolumn{1}{c}{\Tl} 	& \multicolumn{1}{c}{\Th} 	& \multicolumn{1}{c}{\Tl} 	 \\ \hline
10					& 	20	&  2.0	& 	14	&  0.5       \\ 
20 					& 	50	&  2.0	&  	20	&  0.5	\\ 
30					& 	60	&  2.0	& 	24	&  0.5	\\ 
40 					& 	65	&  2.5	&  	28    &  1.0	\\ 
50					& 	75	&  3.0	& 	30    &  1.0	\\ 
60 					& 	75	&  3.0	& 	34    &  1.0	\\ 
70					& 	60	&  3.5	& 	36    &  1.0	\\ 
80 					& 	65	&  3.5	& 	38    &  1.0	\\ 
90 					& 	70	&  4.0	& 	40    &  1.0	\\ 
\hline
\hline
 \end{tabular*}
\caption{High (\Th) and low (\Tl) temperatures used in the DA and SA runs of 
spin-glass problems with different densities. The temperature values used in 
our simulations are unitless.}
\label{table:diff_density_final_parameters}
\end{table}

\newpage
\bibliographystyle{apsrevtitle}
\bibliography{ref}

\begin{thebibliography}{72}
\bibitem[{pre()}]{precision}
\bibinfo{note}{{We have used $10^{-5}$ as the precision in our numerical
  simulation.}}
\end{thebibliography}

\end{document}